\documentclass{article}

\usepackage[english]{babel}

\usepackage[letterpaper,top=2cm,bottom=2cm,left=3cm,right=3cm,marginparwidth=1.75cm]{geometry}

\usepackage{amsmath}
\usepackage{graphicx}
\usepackage{float}
\usepackage[colorlinks=true, allcolors=blue]{hyperref}

\usepackage[natbibapa,nodoi]{apacite}
\title{Don’t Let Your Likert Scales Grow Up To Be Visual Analog Scales: Understanding the Relationship Between Number of Response Categories and Measurement Error}

\usepackage{authblk} 
\author[1]{Siqi Sun}
\author[1]{Karen M. Schmidt}
\author[1,2]{Teague R. Henry}

\affil[1]{Department of Psychology, University of Virginia}
\affil[2]{School of Data Science, University of Virginia}

\date{} 

\begin{document}

\maketitle

\begin{abstract}
The use of Visual Analog Scales (VAS), which can be broadly conceptualized as items where the response scale is 0-100, has surged recently due to the convenience of digital assessments. However, there is no consensus as to whether the use of VAS scales is optimal in a measurement sense. Put differently, in the 90+ years since Likert introduced his eponymous scale, the field does not know how to determine the optimal number of response options for a given item. In the current work, we investigate the optimal number of response categories using a series of simulations. We find that when the measurement error of an item is not dependent on the number of response categories, there is no true optimum; rather, reliability increases with number of response options and then plateaus. However, under the more realistic assumption that the measurement error of an item increases with the number of response categories, we find a clear optimum that depends on the rate of that increase.  If measurement error increases with the number of response categories, then conversion of any Likert scale item to VAS will result in a drastic decrease in reliability. Finally, if researchers do want to change the response scale of a validated measure, they must re-validate the new measure as the measurement error of the scale is likely to change.  
\end{abstract}

\section{Introduction}

The precision and utility of psychological assessments are contingent not only on the construct being measured or the specific item text, but also on the specific response scale employed. Traditionally, Likert scales have dominated psychological research with their fixed point ordinal system, ranging from ``Strongly Disagree'' to ``Strongly Agree'' \citep{likert_technique_1932}. Likert scales consist of a series of statements or items, and respondents are asked to indicate their level of agreement or disagreement with each statement. A traditional 5-point Likert scale has five response options, ranging from "Strongly Disagree" to "Strongly Agree", with a neutral midpoint such as "Neutral" or "Neither Agree nor Disagree". The respondents select the option that best represents their point of view on the given statement, and then the responses are numerically coded for analysis. The conventional use of Likert scales generally involves a relatively small number of response options (2-7 points), resulting in limited variability and subsequent challenges in data analysis \citep{sung_visual_2018}. Despite its popular use, the limited variability of Likert scales to accurately model and capture the nuances of human feelings remains heavily questioned.

As an alternative to Likert scales, researchers often use Visual Analog Scales (VAS), particularly in computerized Ecological Momentary Assessment (EMA) designs \citep[e.g.][]{zealley_growing_1969, hayes_experimental_1921, bosch_measurement_2019, sung_visual_2018,flynn_comparison_2004,guyatt_comparison_1987,jaeschke_comparison_1990}. A traditional VAS (such as one administered in a pen-and-paper instrument) has a horizontal line anchored by opposing extremes (e.g., "not at all stressed" to "extremely stressed"), where respondents mark a point that reflects their perception on the issue being measured, and the distance from the line's beginning to the mark represents the quantitative score. This quantitative score is usually coded from 0-100. Electronic versions of VAS items typically use a slider, where respondents move an indicator along a line to mark their response. Several clinical studies have been found to support the reliability of using VAS in pain assessment \citep{bijur_reliability_2001,averbuch_assessment_2004}. Previous work has examined reliability and validity for a number of psychological constructs using VAS type scales, including anxiety \citep{williams_psychometric_2010,davey_one-item_2007,abend_reliability_2014} and stress \citep{lesage_clinical_2012}. VASs are increasingly being used in combination with EMA settings because of the convenience they offer to capture real-time, in-the-moment data. In EMA studies, where frequent, repeated measurements are required throughout the day, the use of brief, simple scales is crucial to reduce participant burden and ensure high response rates \citep{allen_single_2022}. As a result, single-item measurements are the most common in EMA studies, allowing efficient data collection without overwhelming participants. 

Methodological research suggests that there is an optimal number of measurement categories for any given question, and this number is dependent upon the type of scale and respondents' motivational and cognitive characteristics \citep{ferrando_kernel_2003,scherpenzeel_validity_1997}. Factors that can impact the optimal number of measurement categories include method effect, which is defined as systematic variance attributable to the measurement method rather than to the constructs the measures represent \citep{podsakoff_common_2003}, and response style, which is defined as respondent's tendency to react systematically to measurement items independent of the nature of the measurement \citep{paulhus_measurement_1991}. Likert scales and VAS scales can be viewed as the same type of ordinal scale with differing numbers of response categories. Under this perspective, a VAS scale is simply a 101 (or 100) point Likert scale. The question then is: What is the appropriate number of response categories for a given item? There are only three possible answers to this question: There either is an optimal number of response options for any given item that serve to maximize its information content regarding a given construct, or that more response options are always better, or increasing the number of response options will have no impact at all. To illustrate this issue with a simple example. Consider an item that measures negative mood on a 5-point Likert scale. We might reasonably assume that a participant can, with a relatively high degree of accuracy, discriminate between a value of 1 (very much like me) and 2 (somewhat like me). However, will that participant be able to discriminate between 1 and 2 on a 20 point scale with the same degree of accuracy as in the 5 point scale?  How about a 100 point scale, or a 1000 point scale? At a certain point, the respondent will no longer be able to make meaningful distinctions between adjacent response options. This thought experiment very reasonably implies that the measurement error of an item (which has a 1-to-1 relation to the ability of a respondent to discriminate between adjacent options, as we will establish below) will be positively related to the number of response options.  In the following work, we explore the consequences of this relation via Monte Carlo simulation. 

Researchers hold a variety of views on how to determine the appropriate number of response categories. Many researchers suggest that increasing the number of response categories leads to higher reliability \citep{alwin_reliability_1991,alwin_information_1992} and higher validity \citep{alwin_information_1992,alwin_feeling_1997,muniz_item_2005,preston_optimal_2000,thomas_how_2004}. \citet{edward_f__krieg_biases_1999} mathematically demonstrated biases induced by coarse measurement, arguing that the more points the better, with a continuous scale being the optimal choice; Krieg encourages the use of graphic scales (like the VAS) to avoid the biases introduced by coarse measurement. Proponents of finer-grained scales argue that restricting the number of categories constrains respondents' ability to precisely convey their feelings \citep{viswanathan_does_1996}. \citet{champney_optimal_1939} found that the commonly used five- or seven-point scales often resulted in a significant loss of reliability and suggested using an 18- to 24-point measurement scale because their empirical data showed that increased refinement up to about 22 points could achieve higher reliability without adding excessive random error, thus optimizing the balance between added significant discrimination and random error. 

Conversely, a considerable amount of research suggests that a relatively small number of response categories should be used. Supporters of 7-point Likert scales argue that 7 is the amount of information humans are able to maintain in their immediate memory span \citep[e.g.][]{miller_magical_1956}. Multiple studies have found that 5-point Likert scales are able to produce satisfactory reliability \citep[e.g.][]{preston_optimal_2000,weng_scale_1998}; additional categories beyond 5-point increase reliability, but the gain is minimal and may plateau as the number of categories increases \citep{loevinger_attenuation_1954}. \cite{lozano_effect_2008} examined the effect of the number of response categories on the reliability and validity of Likert-type rating scales using simulated data and found that reliability and factorial validity improved as the number of response categories increased, with the optimal range being 4 to 7 categories; fewer than four categories reduced psychometric quality, while gains were minimal beyond seven categories. \citet{mckelvie_graphic_1978} argues that 5 or 6 response categories should be used for the ease of coding and scoring without sacrificing reliability. \citet{lei_chang_psychometric_1994} showed that the 6-point scale had lower reliability compared to the performance of the 4-point scale. \citet{preston_optimal_2000} found that reliability tends to decline in scales with more than 10 response categories, with 7-point scale exhibiting the best psychometric properties. \cite{van_laerhoven_comparison_2004} found that children preferred Likert scales and rated them as easiest to complete, while simple and numeric VAS showed comparable reliability but were less favored. \citet{adelson_measuring_2010} found that 3rd and 6th grade students were able to discriminate 5-point scales but prefer 4-point scales compared with the 5-point response option. Additionally, \citet{alan_effect_2020} demonstrated that for younger populations, scales with three and four response categories showed superior psychometric properties, including increased reliability and improved model-data fit, as compared to two-category scales. \citet{lee_search_2014} found that 4-, 5-, and 6-point scales display nearly identical psychometric properties but deteriorate when reducing number of categories from 3 to 2. In a more recent empirical study, \citet{simms_does_2019} showed that the VAS form of the commonly used Big Five Inventory \citep{john_big-five_1999} was inferior to the 6 option Likert version of the scale across several different validity measures. \citet{schmidt_more_2010} studying chronic pain patients showed that rescoring lengthy rating scales by reducing the number of response categories led to improved measurement properties. 

While substantial effort has been devoted to finding the optimal number of response categories, nevertheless, low consensus has been achieved throughout a long history of research. Regardless of the number of response categories researchers choose for Likert scales, there seem to be serious issues in the subsequent analysis, including measurement errors induced by the ordinal nature of data, difficulties in determining an optimal number of response categories due to conflicting research, and limitations in respondents' ability to accurately convey their feelings with a restricted number of categories \citep{sung_visual_2018}. Proponents of Visual Analog Scales (VASs), commonly administered as 100-point scales, argue that the measurement performance of VASs benefits from having finer granularity; additionally, VASs can be treated as continuous measurement without worrying about finding appropriate statistical methods to analyze ordinal data.

However, we question whether humans are able to distinguish fine intervals between adjacent categories on a scale with a high number of response options. Is more measurement variance better for measurement performance? In other words, does the added variance help statistical performance, or is this additional variance simply additional error? So far, we do not have a tool to precisely answer this question. Previous literature investigating this issue often assesses questionnaire data with multiple items using an increasing number of response categories, showcasing that increasing the number of categories improves performance up to a certain point. However, when operationalizing their studies, researchers often use common factor models which control for measurement error in the model instead of independently assessing the impact of the error on measurement performance. Furthermore, empirical work such as \cite{preston_optimal_2000}, \cite{adelson_measuring_2010}, and \cite{lei_chang_psychometric_1994} have often suggested that there is an optimal number of response categories for a given set of items (which differs based on the specific scale and sample), while simulation studies such as \cite{wu_can_2017}, \cite{edward_f__krieg_biases_1999}, and \cite{lee_search_2014} have demonstrated that small numbers of response categories (2-4) are generally sub-optimal, while increasing the number of response options either leads to an increase in reliability/validity or a increase followed by a plateau. 

In this paper we present an exploration of the optimal number of response categories from a psychometric perspective. We use Monte Carlo simulation to evaluate how different factors, such as number of items, sample size, predictive strength, and most importantly measurement error change the impact of the number of response categories on several different measures of psychometric reliability. Here, we use a specific definition of measurement error, the item variable construction, that allows us to intuitively define it independently of the number of response categories.

\subsection{Item Variable Construction of the GRM Model}

Within psychometrics, Item Response Theory (IRT) is a widely used framework for scaling both persons and items, often on dichotomously scored or polytomously scored sets of responses. One consideration that makes it difficult to study the relation between number of response categories and measurement error is that in IRT models, the parameterization of measurement error is closely tied to the IRT framework. Specifically, in 2PL models and the multi-category Graded Response Model, measurement error is parameterized as the discrimination of the item, which can be conceptualized as how easy it is for a respondent to move between adjacent categories as their underlying latent variable changes values. While there is a one to one relation between measurement error and the discrimination parameter, it can be difficult to compare how different values of discrimination correspond to different amounts of measurement error. Fortunately, as is so often the case in psychometrics, \citet{lord_statistical_1968} provide a framework for defining the measurement error of an item that is wholly independent of the specific measurement model being used with their \textit{item variable} construction. In their original work, the item variable construction was used to provide justification for the ogive construction of the item characteristic curve. This item variable construction can be thought of as a general item location-scale measurement model. Conceptually, this is a general model for the measurement properties of a specific item (ala, a given item text), before we consider the number of available response options. Here, we use the item variable construct to reparameterize \citet{samejima_estimation_1969}'s Graded Response Model.

Denote a fully continuous measurement process as the following:
$$\theta_i \sim N(0,1)$$ as our true latent score, while our item variable is defined as:
$$\gamma_{ij} \sim  N(\theta_i, \sigma_j)$$
where $\sigma_j$ is the \textit{theoretical} continuous measurement error variance for item $j$. Conceptually, $\sigma_j$ can be thought of as the true measurement error of the item with respect to the underlying construct, ala, how accurate is the item at measuring the construct. Analytically, there is a one to one relationship between $\sigma_j$ and the discrimination parameter of the graded response model.

 Let us assume that the response option chosen $y_{ij}=k$ occurs when the $\gamma_{ij}$ value falls between specific threshold values ($\beta_{j,k-1} -\beta_{j,k}$) with probability 1. As we are integrating over the item variable, this leads to the ogive version of the Graded Response Model.

The probability in responding in the $k$th response option conditional on the true score $\theta_i$ is then:
$$ P_{ik}(\theta_i)=p(\tilde{y}_{ij} = k | \theta_i,  \sigma_j) = \frac{1}{\sigma_j\sqrt{2\pi}} \int_{\beta_{j,k-1}}^{\beta_{j,k}} \exp -\frac{1}{2}\left(\frac{\gamma_{ij} - (\theta_i )}{\sigma_j}\right)^2 d\gamma_{ij}$$The item characteristic curve for response option k (ala, the probability one responds greater than k) is then:

$$ICC_k(\theta_i)=p(\tilde{y}_{ij} > k | \theta_i,  \sigma_j) = \frac{1}{\sigma_j\sqrt{2\pi}} \int_{\beta_{j,k}}^{\infty} \exp -\frac{1}{2}\left(\frac{\gamma_{ij} - (\theta_i)}{\sigma_j}\right)^2 d\gamma_{ij}$$

Note that this is the item response characteristic curve for a ogive graded response model, as per \citet{samejima_estimation_1969}. What this construction allows us to do is define measurement error on the same scale of the underlying construct. We in turn use this construction to generate data in our simulation study below.

\section{Methods}

\subsection{Data Generation Process}
In this simulation study, we employed the aforementioned item variable construction of the graded response model to generate item responses while systematically varying measurement error, number of response categories, number of items, and sample size. To generate our data, we first sampled participant-level latent constructs from a standard normal distribution (as is commonly used in IRT). These latent construct values represent the true scores against which we compare estimates derived from the measured items. Subsequently, we utilized the item variable construction of the Graded Response Model (GRM) previously described to simulate item responses based on these true scores.

We set the threshold values $\beta$ in the following way. For each number of response categories, a sequence of evenly spaced values between -2 and 2 was generated. This sequence included one more value than the number of response categories; the first and last values of this sequence (corresponding to -2 and 2) were excluded, leaving a set of thresholds that correspond to the boundaries between the response categories. These thresholds were designed to evenly partition the interval from -2 to 2 into number of response categories segments, which covers approximately 95\% of the responses in a standard normal distribution. For dichotomous scales, a single threshold value of 0 was used. This corresponds to a standard binary split of the latent trait distribution.

This construction of the threshold parameters ensured that each item measured the same interval of the underlying latent construct while systematically varying the number of response options. By maintaining consistent intervals between threshold values across different item formats, we controlled for item difficulty and focused solely on the influence of the number of response categories on measurement precision. 

This study compared conditions with one and three items to investigate the impact of number of items on measurement performance. Our interest lies in Ecological Momentary Assessment (EMA) designs where the use of visual analog scales are most common. EMA designs often prioritize brevity and participant burden reduction, which often employ a limited number of items, sometimes even a single item, in contrast to traditional survey research. Given the frequent and often intrusive nature of EMA data collection, using single-item measures can significantly decrease participant fatigue and improve compliance. In our single item condition, the observed score was used in subsequent analyses, while in the three-item condition, an average of the observed scores was used.

In addition to our interest in evaluating the recovery of the underlying latent variable, we were also interested in how changing the number of response categories impacts statistical inference. To evaluate this, we simulated a third variable that was a linear function of the underlying latent variable, and evaluated the estimated regression coefficient and, more importantly, the standard error of that regression coefficient. To simulate this third variable, we generated a predictor variable by adding random noise to a linear function of the true latent construct. Specifically, we created the predictor by multiplying the true latent score by a fixed coefficient (0.5) and then adding normally distributed random noise with a standard deviation of 0.2. This approach allowed us to examine the impact of measurement error on parameter estimation and inferential statistics. As our interest was purely in how the properties of the measured items impact inference, we did not vary the residual variance or true regression coefficient.

In the aforementioned conditions, we assumed that measurement error was independent of the number of response categories. To further investigate conditions where measurement error is dependent on the number of response categories, we introduced three additional conditions in which measurement error exhibited monotonically increasing linear relationships with the number of response categories. 

\subsection{Simulation Parameters}
To compare the performance of ordinal Likert scales and VAS scales with differing number of response options, we simulated individualized items using the ogive version of the graded response model construction fully crossing the following four factors:

\begin{itemize}
    \item Number of response categories, which are integers from 2 to 100, representing different types of scales, from binary (two categories) to continuous-like scales with 100 categories.
    \item Measurement error of the item $\sigma_j$. The sequence ranged from 0.1 to 1.0, with increments of 0.1, resulting in a set of 10 standard deviation values.
    \item Number of items (single item or three items).
    \item Sample size (100, 500, 1000).
\end{itemize}

For conditions where measurement error was dependent on the number of response categories, we adjusted the error sequence to create small, medium, and large linear dependencies. Specifically, the measurement error was mapped directly to the number of response categories, which ranged from 2 to 20, in a one-to-one relationship. In the small dependency condition, measurement error ranged from 0.05 to 0.5, increasing in increments of 0.025. In the medium dependency condition, the error values ranged from 0.1 to 1.0, with increments of 0.05. In the large dependency condition, the error values ranged from 0.2 to 2.0, with increments of 0.1. Note: these categorizations of small, medium and large dependency are relative to the scaling of the underlying latent variable (i.e. the latent variable has a standard deviation of 1, so in the large dependency example, measurement error maxes out at twice the scale of the latent variable), and should not be taken as standard categorizations for future use. 

\subsection{Outcomes}
This study assessed measurement properties using two primary outcomes: (a) recovery rate of the true score, calculated as Spearman's correlation between true and item response scores, and (b) standard error (SE) of the estimated regression coefficient. The SE of the estimated regression coefficient was obtained, in the standard fashion, as the standard error of the slope coefficient from a linear regression model predicting standardized observed responses (aggregated row means) from a standardized predictor variable, which included measurement error. To examine changes in these outcomes across different conditions, we calculated delta values (difference between consecutive data points) and visualized them in delta plots (Figures \ref{fig:deltarecov} and \ref{fig:deltase}). While the bias of the regression coefficient was also explored, it yielded no substantial findings beyond the patterns observed with Spearman's correlation and is therefore detailed in the Supplementary Materials.

\section{Results}

\subsection{When Measurement Error is Independent of Number of Response Categories}
\begin{figure}[H]
    \centering
    \includegraphics[width=.99\textwidth]{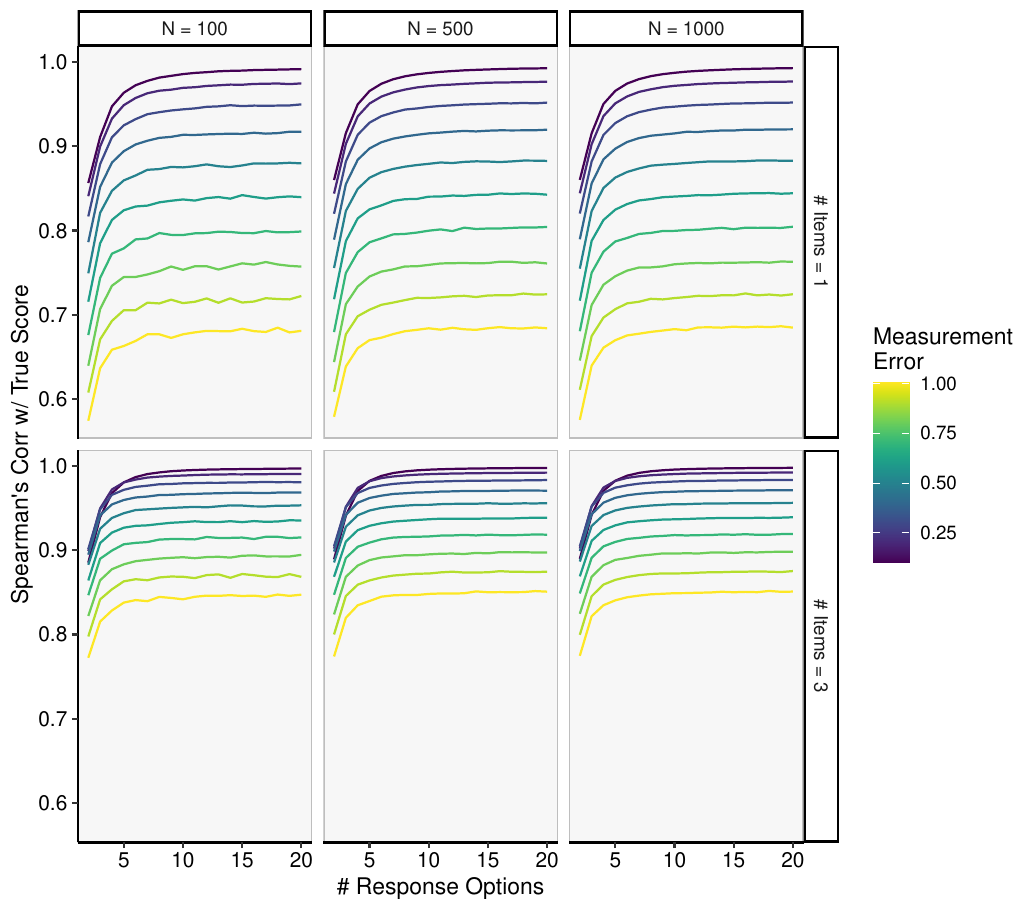}
    \caption{True - Estimated Spearman's Correlations. This plot illustrates the estimated Spearman's correlations between the true score and the observed score across varying numbers of response options. The colored lines represent the estimated Spearman's correlation values, with the gradient from yellow to purple indicating increasing levels of measurement error (yellow for higher error and purple for lower error). The row facets correspond to the number of items included in the measurement (1 item in the top row, 3 items in the bottom row), while the column facets represent different sample sizes (N = 100, 500, 1000).}
    \label{fig:recov}
\end{figure}

 Figure \ref{fig:recov} depicts the relationship between the number of response options, the amount of measurement error and the recovery of true latent scores, analyzed across three sample sizes (100, 500, and 1000). These results suggest that as the number of response options increases, the recovery of the true latent scores improves up to a point, then it plateaus. Our results demonstrate that increasing the number of items noticeably improved measurement performance, particularly for items with higher measurement error. This finding is consistent with well-established principles in psychometrics, where adding more items generally enhances the reliability and precision of measurement instruments. The observed improvement was especially pronounced in scenarios involving higher measurement error, a pattern that aligns with expectations. Sample size did not significantly affect measurement performance. We observed a trend of diminishing returns, where the benefit of adding more response options became minimal after a certain point. Importantly, our findings suggest that increasing the number of response options does not compromise measurement performance, but nor does it improve measurement performance after a certain number of response categories.

\begin{figure}[H]
    \centering
    \includegraphics[width=.99\textwidth]{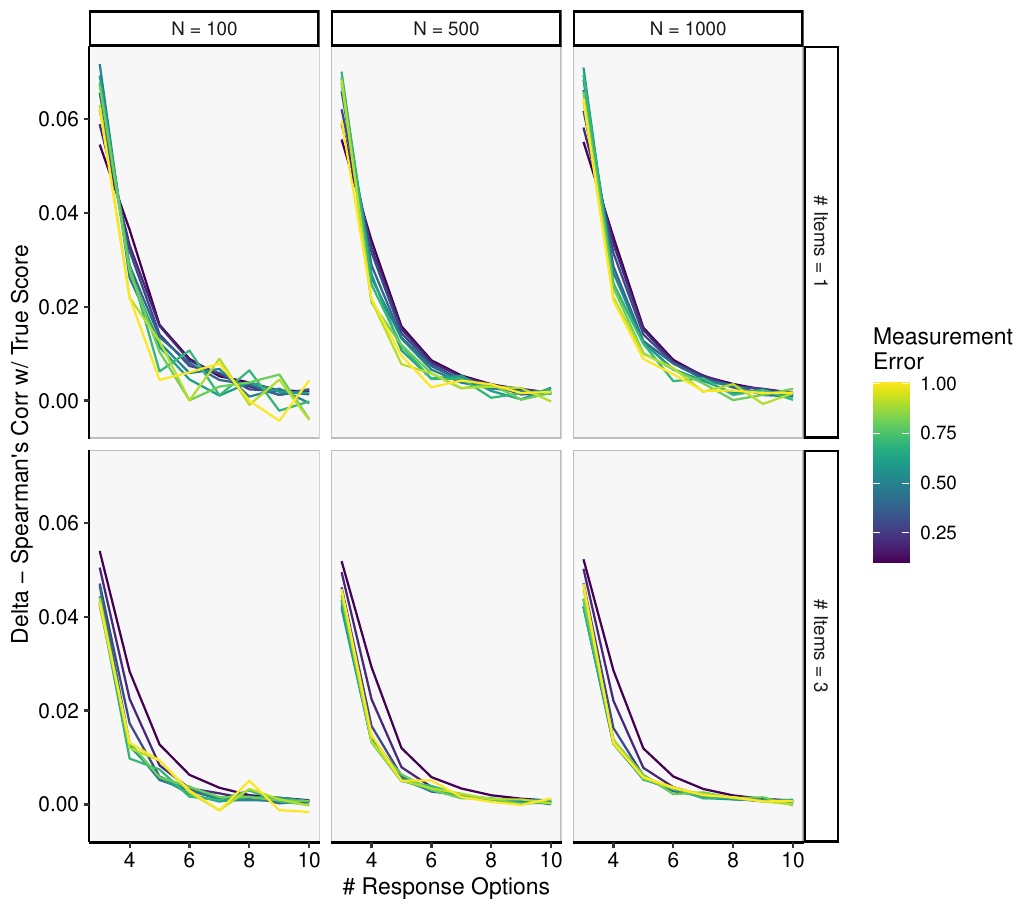}
    \caption{Delta - Spearman's Correlations with True Score. This plot illustrates the changes in Spearman's correlation between the true score and the observed score as the number of response options increases. The colored lines represent the delta values, with the gradient from yellow to purple indicating increasing levels of measurement error (yellow for higher error and purple for lower error). The row facets correspond to the number of items included in the measurement (1 item in the top row, 3 items in the bottom row), while the column facets represent different sample sizes (N = 100, 500, 1000).}
    \label{fig:deltarecov}
\end{figure}

Figure \ref{fig:deltarecov} explores the change in Spearman's correlation as the number of response options increases, analyzed across the same three sample sizes and number of items as in Figure 1. The delta plots offer a different perspective on the same data by quantifying how much the Spearman's correlation improves as additional response options are introduced, allowing us to visualize any differences in the rates of improvement in latent score recovery. Notably, the steepest improvements occur when moving from very few response options to a moderate number (approximately 4-6), after which the gains begin to diminish. This trend is particularly evident in scenarios with higher measurement error, where the delta values start higher and decrease more sharply. This suggests that increasing the number of response options is most beneficial when initial response options are limited and measurement error is high. As with the raw Spearman's correlations, the impact of sample size on delta values is minimal.

\begin{figure}[H]
    \centering
    \includegraphics[width=.99\textwidth]{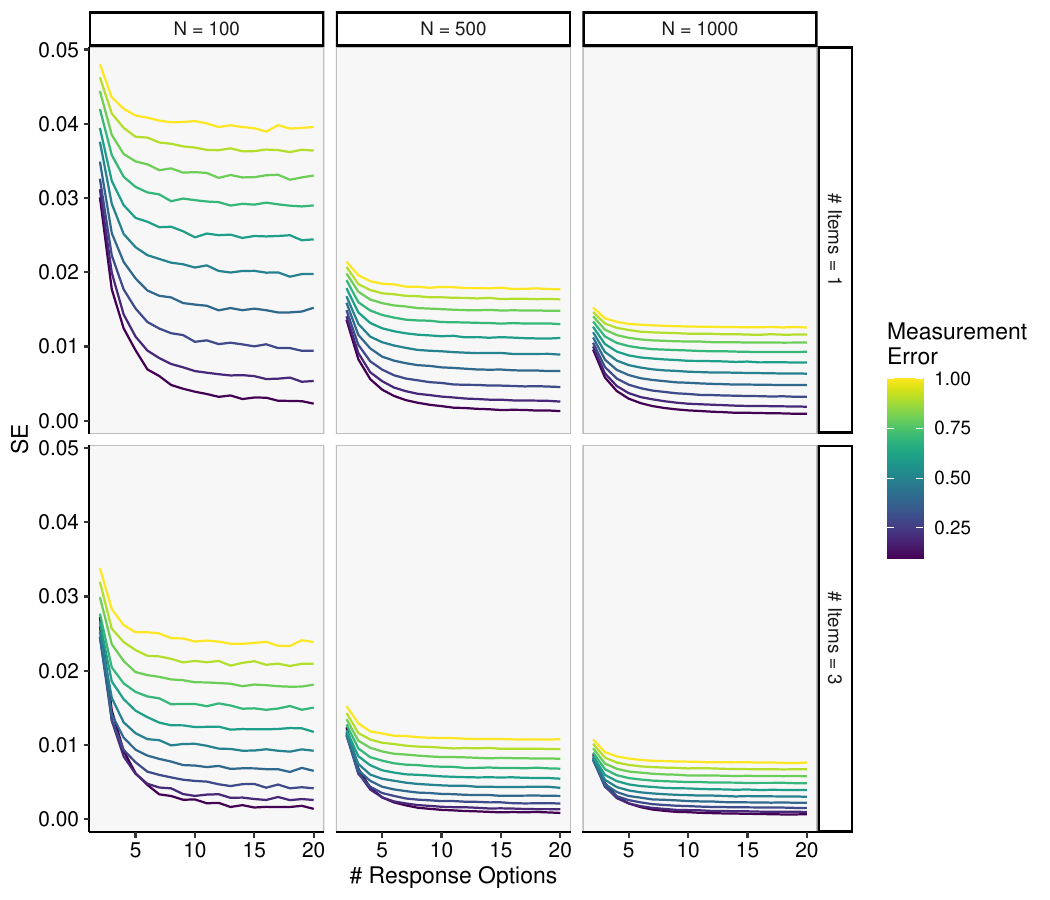}
    \caption{Standard Error of The Estimated Regression Coefficient. This plot shows the standard error (SE) of the estimated regression coefficient across varying numbers of response options. The lines in the plot represent the SE values, with colors transitioning from yellow to purple to indicate increasing levels of measurement error (yellow for higher error and purple for lower error). The row facets correspond to the number of items included in the measurement (1 item in the top row, 3 items in the bottom row), and the column facets represent different sample sizes (N = 100, 500, 1000).}
    \label{fig:regse}
\end{figure}

Figure \ref{fig:regse} presents the standard error (SE) of the estimated regression coefficient as a function of the number of response options, across varying levels of measurement error, sample sizes, and number of items. Increasing the number of response options generally reduces the bias in SE, with a more pronounced effect observed in smaller sample sizes (N = 100). This reduction in bias with more response options is particularly significant when measurement error is high. As sample size increases, the SE becomes small, which is entirely expected as larger samples provide more accurate and stable estimates as a general rule. The number of items also plays a crucial role; using three items instead of one consistently reduces the bias across all conditions. This finding aligns with the well-known principle that using more items can improve the accuracy and reliability of statistical estimates. Again, like with the previously described results on the recovery of the true latent score, these results show that, regardless of other properties, increasing the number of response categories reduces the estimate of the SE up to a point (6-8 response categories), after which the effect of increasing response categories is minimal.

\begin{figure}[H]
    \centering
    \includegraphics[width=.99\textwidth]{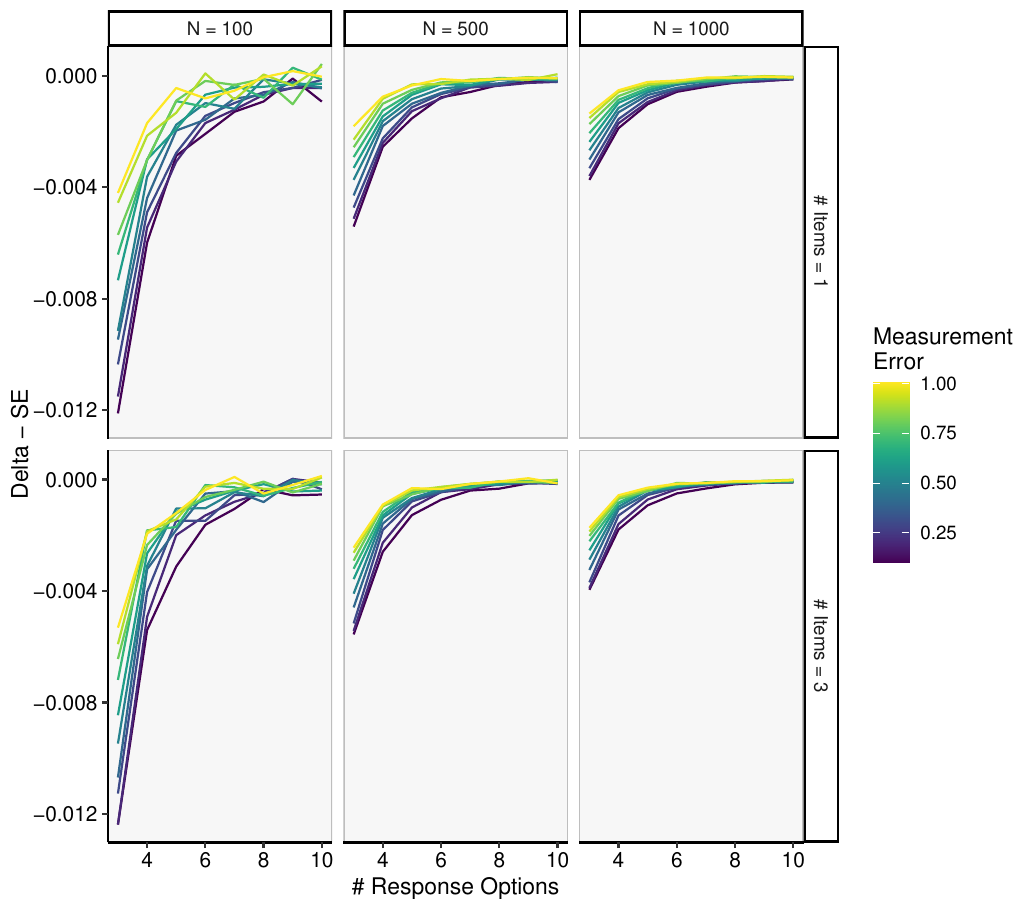}
    \caption{Delta - Standard Errors. This plot illustrates the change in the standard error (SE) of the regression coefficient as the number of response options increases. The lines represent the delta of SE, with colors indicating varying levels of measurement error (yellow for higher error and purple for lower error). Row facets indicate the number of items in the measurement (1 item in the top row, 3 items in the bottom row), while column facets represent different sample sizes (N = 100, 500, 1000).}
    \label{fig:deltase}
\end{figure}

Figure \ref{fig:deltase} presents the delta values for the standard error as the number of response options increases. The delta values quantify the change in SE with each additional response option. As observed in Figure 3, the most substantial reductions in bias occur when the number of response options is initially low, especially in conditions with high measurement error and smaller sample sizes. The delta plots show that as the number of response options increases, the bias in SE reduces significantly, particularly for smaller sample sizes. However, beyond a certain point (approximately 6-8 response options), the rate of bias reduction flattens, indicating diminishing returns. This pattern is consistent across all sample sizes and item numbers, underscoring that while adding more response options can significantly reduce SE, the benefits plateau after a moderate number of categories. These results tie back to the findings in Figure 3, where larger sample sizes and more items consistently lead to lower bias and more reliable estimates.

\subsubsection{Bias of the Regression Coefficient}

We examined the bias of the regression coefficient, and found results that were wholly consistent with what is shown in Figures 1-4. As such, we included these figures in the Supplementary Materials, and provide a brief description here. The bias of the regression coefficient for the third variable follows a trend similar to the recovery of the true score as presented in Figure \ref{fig:recov}: Both benefit from a higher number of response options, lower measurement error, and the use of more items, while not benefiting from an increase in sample size. Similar to Figure \ref{fig:regse}, the bias of the standard error in the regression relationship to the third variable benefits from a higher number of response options, lower measurement error, the use of more items, and a larger sample size. 

In the case where measurement error is independent of the number of response categories, our findings highlight the importance of several key factors in enhancing the accuracy of measurement instruments. We observed that increasing the number of response options generally improves the recovery of true latent scores, as well as reduces bias in the regression coefficient and standard error, particularly when the number of response options reaches a certain threshold (somewhere between 5-10 response options), beyond which the benefits plateau. The inclusion of multiple items significantly bolstered measurement performance, especially in scenarios involving higher measurement error, reinforcing well-established psychometric principles. While sample size did not notably impact the recovery of true latent scores or the bias of the regression coefficient in any analysis condition, it did play a crucial role in reducing the bias of the standard error. While these findings appear to suggest that there is no optimal number of response categories, or more accurately, after a certain number of response categories, adding additional response categories doesn't improve or degrade measurement performance, it is important to emphasize that in these simulations, the measurement error is wholly independent with the number of response categories. While there are certainly theoretical arguments why this assumption might hold (i.e. the measurement error of an item is related to the content of the item rather than the specific response scale), these findings do not agree with the various empirical studies that have shown degraded measurement when the number of response categories becomes too high. We do note that this plateauing behavior has been demonstrated in a number of other simulation studies that assess measurement with differing numbers of response categories. To assess one possible cause of this behavior, in the next section we present simulation conditions where the number of response categories directly impacts the measurement error of an item. Here, this corresponds to respondents being increasingly unable to distinguish between adjacent response categories as the possible options becomes high.

\subsection{When Measurement Error is Dependent on Number of Response Categories}

\begin{figure}[H]
    \centering
    \includegraphics[width=.99\textwidth]{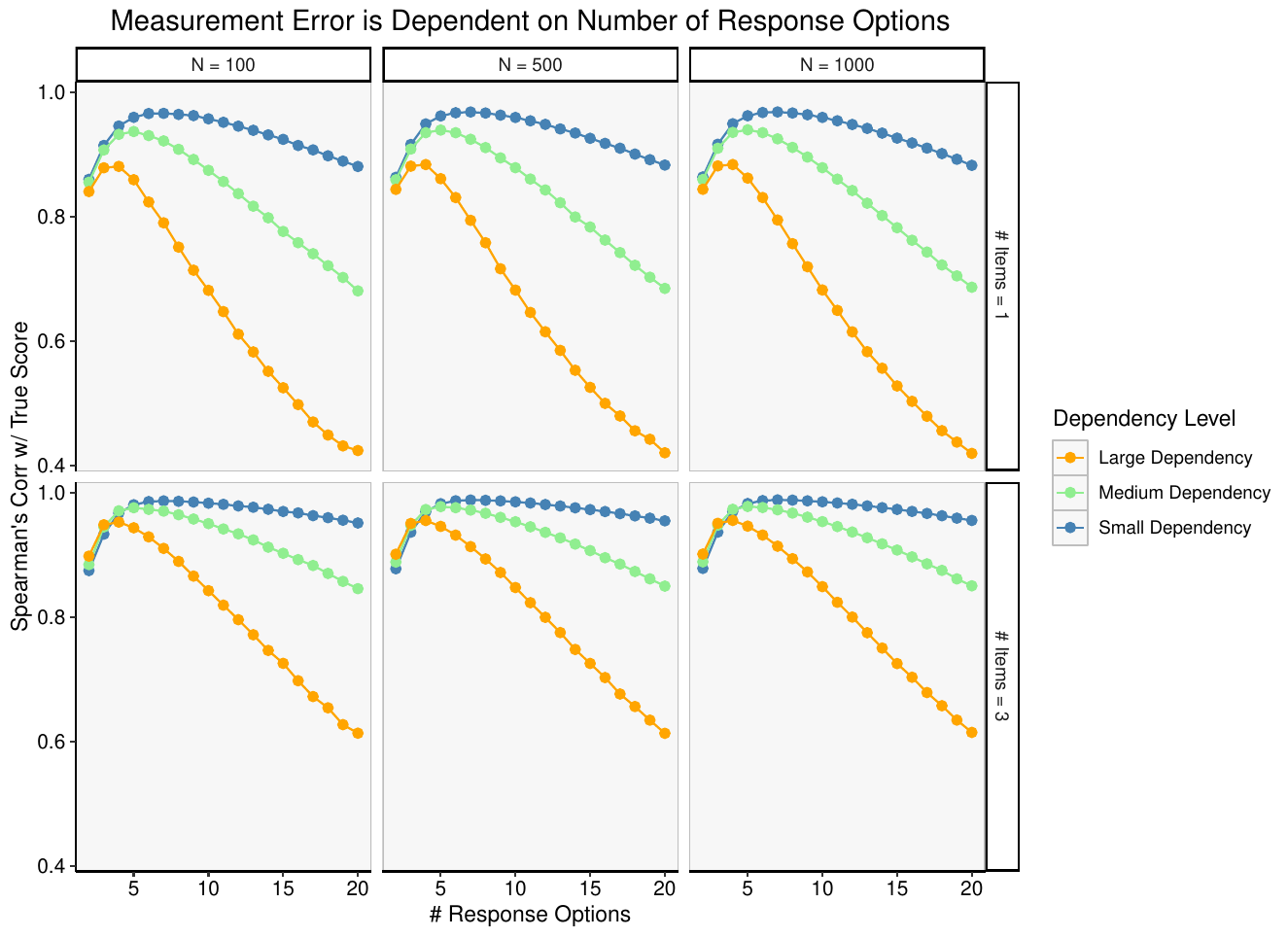}
    \caption{True - Estimated Spearman's Correlations When Measurement Errors Are Dependent On Categories. This plot illustrates the estimated Spearman's correlations between the true score and the observed score across varying numbers of response options and dependency levels. The colored lines represent the estimated Spearman's correlation values, with blue indicating small measurement error dependency, green for medium dependency, and orange for large dependency. The row facets correspond to the number of items included in the measurement (1 item in the top row, 3 items in the bottom row), while the column facets represent different sample sizes (N = 100, 500, 1000).}
    \label{fig:combinedrecov}
\end{figure}

Figure \ref{fig:combinedrecov} demonstrates the relationship between the number of response options, the dependency of measurement error on the number of response categories, and the recovery of true latent scores. The analysis is conducted across three sample sizes (100, 500, and 1000) and two item conditions (1 item versus 3 items). When measurement error is dependent on the number of response categories, the recovery of true latent scores declines steadily as the number of response options increases, especially under conditions of higher dependency (orange). In scenarios with small dependency (blue), the optimal recovery is observed at 7 response options, after which the recovery rate plateaus. Medium dependency (green) demonstrates a similar pattern, with optimal recovery at 5 response options, but the overall recovery is lower than that observed under small dependency conditions. For large dependency (orange), the recovery declines more quickly as the number of response options increases, with an optimal number of response options at 4, after which the performance drops significantly. This suggests that under high dependency, adding too many response categories exacerbates measurement error, leading to diminished recovery rates. Notably, increasing the number of items (as shown in the bottom row) significantly improves recovery, particularly for large dependency levels. This effect is consistent across all sample sizes, indicating that including more items helps stabilize measurement performance even when measurement error is highly dependent on the number of response options. Interestingly, sample size had a relatively minor impact on recovery performance. While larger sample sizes provided slightly better recovery overall, the main effects were driven by the number of response categories and the degree of measurement error dependency.

\begin{figure}[H]
    \centering
    \includegraphics[width=.99\textwidth]{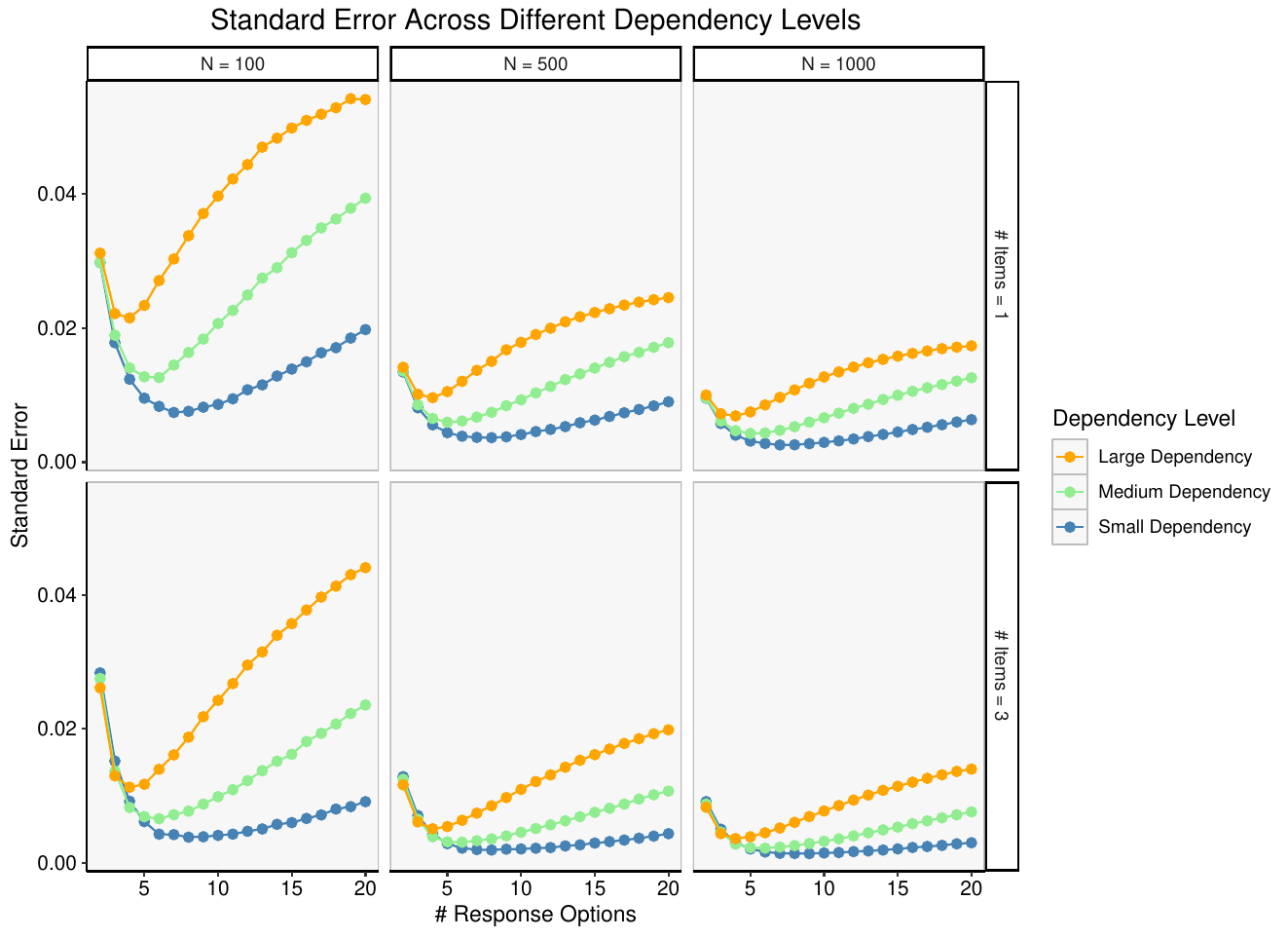}
    \caption{Standard Error of The Estimated Regression Coefficient When Measurement Errors Are Dependent On Categories. This plot shows the standard error (SE) of the estimated regression coefficient across varying numbers of response options and dependency levels. The lines in the plot represent the SE values, with blue indicating small measurement error dependency, green for medium dependency, and orange for large dependency. The row facets correspond to the number of items included in the measurement (1 item in the top row, 3 items in the bottom row), and the column facets represent different sample sizes (N = 100, 500, 1000).}
    \label{fig:combinedse}
\end{figure}

Figure \ref{fig:combinedse} illustrates the standard error (SE) of the estimated regression coefficient across different numbers of response options, varying levels of dependency between measurement error and response options, sample sizes (N = 100, 500, 1000), and the number of items (1 or 3 items). In larger sample sizes, the SE is generally lower and more stable. For large dependency (orange), the SE decreases initially but rises again after approximately five response options, with this increase being more pronounced in smaller samples. Medium dependency (green) follows a similar trend, with SE stabilizing around five to seven response options before slightly increasing. Small dependency (blue) shows the most stable SE across conditions, continuing to decline up to around 10 response options and remaining relatively low afterward, especially in larger sample sizes. The SE decreases substantially when moving from single-item to three-item measures, suggesting that using multiple items reduces the SE in all conditions.

\section{Discussion}

Our study revisits the question of whether there is an optimal point on the measuring scale that maximizes reliability. Our findings suggest that when measurement error is independent of the number of response options, there is a threshold where the benefit of additional response options begins to plateau, typically between 5 and 10 categories, beyond which additional categories offer diminishing returns in terms of reliability. However, when measurement error is dependent on the number of response options, this plateau disappears. We identified the optimal number of response categories at 7, 5, and 4 for small, medium, and large dependency structures, respectively, with reliability declining noticeably once these thresholds are exceeded. 
 
In our study, we explored the impact of several factors—namely, the number of response options, measurement error, sample size, and the number of items—on the recovery of true latent scores and the associated measurement properties. When measurement error is independent of the number of response categories, we found the benefits of increasing response options generally plateauing between 5 and 10 categories, with higher error levels leading to poorer recovery of true scores and increased bias in the regression coefficients and standard errors. When measurement error is dependent on the number of response options, however, the benefits of additional categories diminish more rapidly. Specifically, the optimal number of response options drops to 4, 5, or 7, depending on the level of error dependency, with reliability declining significantly beyond these thresholds. The use of multiple items, as opposed to a single item, consistently improved measurement performance (as expected), particularly in scenarios with higher measurement error. However, the use of multiple items had no effect on the optimal number of response categories for those items.  Interestingly, while sample size had a noticeable effect on reducing the standard error of the regression coefficients, it did not significantly impact the recovery of true scores or the bias of the regression coefficients. Importantly, when measurement error is independent of the number of response categories, exceeding the optimal number of categories does not negatively impact the psychometric properties of the item; however, when measurement error is dependent on the number of response categories, exceeding the optimal number leads to a decline in reliability. Therefore, we strongly recommend validating scales carefully before determining the optimal response format.

As our results have shown that the dependency structure between measurement error and the number of points is critical, it is important to validate scales with different substantive measurements independently. Besides variations from substantive topics of interests, several factors can contribute to greater measurement error as the number of response options increases. A larger number of options might lead to cognitive overload, where participants become overwhelmed by the choices, making it harder for them to accurately select the best option \citep{miller_magical_1956,baddeley_magical_1994,jacob_over_2024}. This cognitive overload can result in response errors, thereby increasing measurement error. Furthermore, as the number of choices increases, it becomes more difficult for participants to distinguish between options, leading to greater uncertainty and potential errors in responses. Responding styles and participant fatigue could also play a role; for example, participants might find it more tiring or cognitively demanding to respond to 100-point scales, which could increase the likelihood of response bias. Response bias may occur when participants develop a tendency to select the same response repeatedly, regardless of the question, thus introducing systematic error into the data \citep{mcgrath_evidence_2010}. Lastly, ceiling and floor effects may emerge: with more response options, participants might reach the ceiling (selecting the highest option for every item) or the floor (selecting the lowest option for every item), reducing variability and limiting measurement accuracy \citep{wang_investigating_2008}. These factors could introduce additional measurement error, potentially offsetting the benefits of having more response options. Further research is needed to fully understand the trade-offs between response option quantity, participant experience, and measurement quality across various contexts.

Recent work by Haslbeck and colleagues (\citeyear{haslbeck_comparing_2024}) provides an interesting empirical counter point to the theoretical work we present here. In their study, they collected ecological momentary assessment data on affective states, six times a day for 14 days, with between subject randomization to either a 7 point Likert scale or VAS. The most important finding, at least in the perspective of measurement performance, was that the correlations between the affective state measures and external psychopathology measures were significantly greater on the whole when affective states were measured using VASs than when they were measured using the 7 point Likert. This result could be attributed to several factors. First of all, the additional variance afforded by VAS scales may enhance their capacity to capture correlational relationships. For example, \cite{kuhlmann_investigating_2017} compared VAS and Likert-type scales in measuring personality traits and found largely equivalent psychometric properties, including reliability, means, and intercorrelations. However, VAS showed a slight advantage for Excitement Seeking by explaining 2.5 percent additional variance in predicting age. \cite{funke_why_2012} studied the performance of semantic differentials using VAS versus 5-point scales and reported that mean correlations measured with VASs were higher than those with 5-point categorical scales. \cite{olsson_polyserial_1982} demonstrated mathematically that the number of response categories influences observed correlations, with lower numbers attenuating the correlation due to reduced precision. This suggests that finer-grained scales, such as VAS, may contribute to better correlation performance by providing greater precision and reducing attenuation. While their model does not account for measurement error, they showed that for 7 response categories, the attenuation factor is approximately 0.945 under mild assumptions, whereas for 100 categories, it approaches near-perfect precision at 0.999. Similarly, \cite{hilbert_influence_2015} demonstrated that while response format influenced psychometric properties, reliability increased with more response categories. Additionally, the within-person nature of EMA studies may mitigate measurement error, particularly in contexts where VAS scales excel at capturing fine-grained, moment-to-moment variations. However, for psychological measurements focused on more trait-like constructs or those less influenced by granularity, this advantage may be less pronounced. Further complicating the picture, we know little about how constructs behave over time, particularly in dynamic, real-world contexts captured by EMA. \cite{averbuch_assessment_2004} found that both VAS and Likert scales showed high average time-series correlations in measuring osteoarthritis pain, however VAS responses showed a wide range for each categorical score with overlaps between categories. \cite{maydeu-olivares_effect_2009} investigated the effect of increasing the number of response alternatives in rating scales using repeated measures designs and compared results across three psychometric models (Classical Test Theory, Item Factor Analysis, and Item Response Theory) and found that increasing the number of response categories significantly improved reliability, particularly in shorter tests, but had a minimal effect on convergent validity and led to poorer model fit for internal structure as the number of categories increased. Further research is needed to understand the balance between measurement error introduced by state measures and that inherent in trait measures, particularly as it relates to scale selection. It is also possible that VAS scales require greater cognitive resources to respond compared to Likert scales, particularly in traditional surveys with lengthy questionnaires. In EMA studies, where a single item is often administered at a time, this is less of an issue because the overall response burden is low. However, in traditional surveys, researchers may need to reduce the number of response options to alleviate the cognitive load for participants, especially in lengthy scales \citep[e.g.][]{schmidt_more_2010}. Participants have limited attention, and reducing the response burden requires a trade-off—either fewer questions or fewer response options. While the within-person nature of EMA reduces measurement error by capturing dynamic responses over time, lengthy surveys with detailed response formats risk increasing error due to participant fatigue or reduced attention. For researchers considering the use of VAS, it is important to balance the advantages of finer granularity with the potential burden introduced by having too many response categories, particularly in studies with repeated measurements or large-scale surveys.
  
We strongly caution researchers against converting established Likert scales to VASs without careful validation. There is a pressing need for a systematic approach to validate VAS, as current empirical knowledge remains limited compared to the abundant evidence available for Likert-type scales. \cite{martinez_developing_2023} highlights that many EMA protocols rely on items originally developed for laboratory or retrospective surveys, which are often adapted without careful validation for momentary assessments. The authors suggest that directly converting existing scales, such as Likert to VAS, without proper validation may compromise measurement accuracy and construct appropriateness. \cite{martinez_validation_2024} discuss the adaptation of scales, noting that most items in their EMA protocol were answered using 7-point Likert scales. The authors explicitly chose Likert scales over VAS due to their ease of use and faster completion times, particularly for younger, older, and less-educated populations. It is essential to recognize that findings from Likert scales cannot be directly applied to VAS due to fundamental differences in their design and interpretation. VAS provides a continuous measure with equidistant points, while Likert scales are ordinal and feature unequal intervals between points. Researchers should only use VASs if the construct they are studying demonstrates an independent structure between measurement error and the number of response options (we believe that this is impossible, but will admit it is an empirically testable question). At the same time, \cite{hasson_validation_2005} showed that while VAS can replace single Likert items for uniform constructs, they are not interchangeable with multi-item Likert indices due to significant differences in absolute levels. The usability of VAS appears to depend on the level of manifest-ness of the construct being measured. Manifest constructs, such as physical sensations or momentary affective states, which are directly observable and continuously varying, tend to align well with VAS. For example, \cite{averbuch_assessment_2004} found that VAS effectively captured physical pain with high reliability and greater granularity compared to Likert scales. Similarly, \cite{haslbeck_comparing_2024} reported that VAS produced stronger correlations with external psychopathology measures when assessing dynamic, momentary affective states in EMA. In contrast, more latent and abstract constructs, such as personality traits or stable psychological characteristics, may not benefit as much from the granularity of VAS. \cite{kuhlmann_investigating_2017} demonstrated largely equivalent psychometric properties between VAS and Likert scales when measuring personality traits, suggesting that finer resolution may add little value for constructs that are less directly observable or less continuous by nature. This highlights the need for further empirical research to determine which constructs are better suited for VAS and under what conditions VAS provides meaningful advantages over traditional Likert scales. Furthermore, our study did not investigate key design aspects of VAS, such as the impact of devices, screen size, orientation, clicking versus dragging, starting position, the use of emoticons, or the effects of different anchor texts and how they interact with the number of response categories \citep{bosch_measurement_2019}. Another common issue is that participants may collapse responses, such as only using multiples of 10 on a 100-point scale \citep{camarda_modelling_2008}. However, in both cases the properties of a VAS scale would result in an increase in measurement error rather than a decrease. Furthermore, there is a need to validate scales with different numbers of response options independently. If measurement error remains consistent regardless of the number of response categories, a single validation may suffice. However, if measurement error varies with the number of response categories, each variation would require separate validation. Importantly, both empirical and methodological research are needed in the future to better understand the psychometric properties of VAS.

One limitation of this study is that we only explored three specific monotonically increasing linear dependency structures between measurement error and the number of response options. More complex nonlinear or multimodal patterns could influence the relationship between response categories and reliability in ways not captured by our simulations. Based on our results, we postulate that as long as the dependency structure is monotonically increasing, an optimal point exists for the number of response options, beyond which reliability decreases. While a completely non-monotonic relationship seems unlikely, future research is needed to examine the effects of nonlinear or multimodal patterns on measurement performance. Another consideration is whether participants perceive a VAS as a truly continuous scale or as a series of implicit categories. While we assume unidimensionality in VAS responses, participants may mentally segment the scale into discrete points, similar to a Likert scale, leading to multimodal response patterns not captured by our current models. Future research should explore how such perceptions influence VAS performance compared to categorical scales. Additionally, VAS may require assumptions such as independence of error and continuous response categories, whereas Likert scales may be more robust to violations of these assumptions due to their structured and discrete nature. Another limitation of this study is the use of composite scores rather than more sophisticated measurement error models, such as Confirmatory Factor Analysis \citep[CFA; e.g.][]{mcardle_latent_2009} or Item Response Theory \citep[IRT; e.g.][]{embretson_item_2000}. However, we are confident in saying that our conclusions would not differ even when using sophisticated modeling approaches. Although these methods can model measurement error and control for it in subsequent analyses, they do not decrease the measurement error in an item. Regardless of analysis method, an item with lower measurement error will always be superior to an item with higher measurement error.  Methods like CFA or IRT are relevant when the number of items increase, rather than when the number of response categories increase. Another limitation of this study is the assumption that the measured construct represents a stable trait. However, it is possible that the measurement of a construct may fluctuate over time, particularly in EMA or longitudinal studies where contextual and temporal factors influence responses. Future research should investigate how temporal variability in the construct being measured interacts with response formats like VAS and Likert scales, as this may have implications for the stability and reliability of the measurements.

\section{Conclusion}
Our findings demonstrate that the optimal number of response options is determined by the dependency structure between measurement error and the number of response categories. When error is independent of the categories, increasing the number of response options enhances measurement reliability, but this improvement plateaus after approximately 10 categories. However, we do not believe the assumption of independent measurement error is reasonable when studying psychometric measurement. When error is dependent on the number of categories, we found that the optimal number of response options is lower, with reliability declining beyond 4 to 7 categories. We note that these optima are highly dependent on our simulation design, particularly with respect to how measurement error increased, and different patterns of dependency will result in different optimal numbers of response options.

That being said, based on these results and considering other studies in this space, we believe that the use of VAS response options for psychological measurement is problematic. While a carefully designed VAS scale—such as one with unidimensionality and minimal options—combined with an appropriate sampling frequency can help mitigate measurement error, these benefits often appear mild and do not outweigh the drawbacks. Importantly, the usability of VAS also depends on the cognitive demands placed on participants, with greater burden occurring in lengthy scales or complex constructs. The benefit of VAS tends to be indistinguishable or small under ideal conditions, yet its drawbacks can be substantial if assumptions such as  independence of error and continuous perception of the scale are violated. Compared to Likert scales, which are inherently more robust to these violations, VAS introduces greater risk when applied without careful consideration.

Fortunately, this is an empirically testable question, and we recommend that future studies systematically validate VAS scales across different psychological constructs, contexts, and design choices, directly comparing them to traditional response formats with 4–7 options. Additionally, we strongly advise against directly transforming an established Likert scale into a VAS format without thorough validation. Moving forward, both empirical and methodological research will be essential to guide researchers in selecting appropriate measurement tools and to better understand the conditions under which VAS may provide meaningful advantages or unintended measurement challenges.

\bibliographystyle{apacite}
\bibliography{Refs}

\begin{thebibliography}{}

\bibitem [\protect \citeauthoryear {%
Abend%
, Dan%
, Maoz%
, Raz%
\BCBL {}\ \BBA {} Bar-Haim%
}{%
Abend%
\ \protect \BOthers {.}}{%
{\protect \APACyear {2014}}%
}]{%
abend_reliability_2014}
\APACinsertmetastar {%
abend_reliability_2014}%
\begin{APACrefauthors}%
Abend, R.%
, Dan, O.%
, Maoz, K.%
, Raz, S.%
\BCBL {}\ \BBA {} Bar-Haim, Y.%
\end{APACrefauthors}%
\unskip\
\newblock
\APACrefYearMonthDay{2014}{}{}.
\newblock
{\BBOQ}\APACrefatitle {Reliability, validity and sensitivity of a computerized visual analog scale measuring state anxiety} {Reliability, validity and sensitivity of a computerized visual analog scale measuring state anxiety}.{\BBCQ}
\newblock
\APACjournalVolNumPages{Journal of behavior therapy and experimental psychiatry}{45}{4}{447--453}.
\PrintBackRefs{\CurrentBib}

\bibitem [\protect \citeauthoryear {%
Adelson%
\ \BBA {} McCoach%
}{%
Adelson%
\ \BBA {} McCoach%
}{%
{\protect \APACyear {2010}}%
}]{%
adelson_measuring_2010}
\APACinsertmetastar {%
adelson_measuring_2010}%
\begin{APACrefauthors}%
Adelson, J\BPBI L.%
\BCBT {}\ \BBA {} McCoach, D\BPBI B.%
\end{APACrefauthors}%
\unskip\
\newblock
\APACrefYearMonthDay{2010}{}{}.
\newblock
{\BBOQ}\APACrefatitle {Measuring the {Mathematical} {Attitudes} of {Elementary} {Students}: {The} {Effects} of a 4-{Point} or 5-{Point} {Likert}-{Type} {Scale}} {Measuring the {Mathematical} {Attitudes} of {Elementary} {Students}: {The} {Effects} of a 4-{Point} or 5-{Point} {Likert}-{Type} {Scale}}.{\BBCQ}
\newblock
\APACjournalVolNumPages{Educational and Psychological Measurement}{70}{5}{796--807}.
\newblock
\begin{APACrefDOI} \doi{10.1177/0013164410366694} \end{APACrefDOI}
\PrintBackRefs{\CurrentBib}

\bibitem [\protect \citeauthoryear {%
Alan%
\ \BBA {} Atalay~Kabasakal%
}{%
Alan%
\ \BBA {} Atalay~Kabasakal%
}{%
{\protect \APACyear {2020}}%
}]{%
alan_effect_2020}
\APACinsertmetastar {%
alan_effect_2020}%
\begin{APACrefauthors}%
Alan, U.%
\BCBT {}\ \BBA {} Atalay~Kabasakal, K.%
\end{APACrefauthors}%
\unskip\
\newblock
\APACrefYearMonthDay{2020}{}{}.
\newblock
{\BBOQ}\APACrefatitle {Effect of number of response options on the psychometric properties of {Likert}-type scales used with children} {Effect of number of response options on the psychometric properties of {Likert}-type scales used with children}.{\BBCQ}
\newblock
\APACjournalVolNumPages{Studies in Educational Evaluation}{66}{}{100895}.
\newblock
\begin{APACrefDOI} \doi{10.1016/j.stueduc.2020.100895} \end{APACrefDOI}
\PrintBackRefs{\CurrentBib}

\bibitem [\protect \citeauthoryear {%
Allen%
, Iliescu%
\BCBL {}\ \BBA {} Greiff%
}{%
Allen%
\ \protect \BOthers {.}}{%
{\protect \APACyear {2022}}%
}]{%
allen_single_2022}
\APACinsertmetastar {%
allen_single_2022}%
\begin{APACrefauthors}%
Allen, M\BPBI S.%
, Iliescu, D.%
\BCBL {}\ \BBA {} Greiff, S.%
\end{APACrefauthors}%
\unskip\
\newblock
\APACrefYearMonthDay{2022}{}{}.
\newblock
{\BBOQ}\APACrefatitle {Single {Item} {Measures} in {Psychological} {Science}} {Single {Item} {Measures} in {Psychological} {Science}}.{\BBCQ}
\newblock
\APACjournalVolNumPages{European Journal of Psychological Assessment}{38}{1}{1--5}.
\newblock
\begin{APACrefDOI} \doi{10.1027/1015-5759/a000699} \end{APACrefDOI}
\PrintBackRefs{\CurrentBib}

\bibitem [\protect \citeauthoryear {%
Alwin%
}{%
Alwin%
}{%
{\protect \APACyear {1992}}%
}]{%
alwin_information_1992}
\APACinsertmetastar {%
alwin_information_1992}%
\begin{APACrefauthors}%
Alwin, D\BPBI F.%
\end{APACrefauthors}%
\unskip\
\newblock
\APACrefYearMonthDay{1992}{}{}.
\newblock
{\BBOQ}\APACrefatitle {Information transmission in the survey interview: {Number} of response categories and the reliability of attitude measurement} {Information transmission in the survey interview: {Number} of response categories and the reliability of attitude measurement}.{\BBCQ}
\newblock
\APACjournalVolNumPages{Sociological methodology}{}{}{83--118}.
\PrintBackRefs{\CurrentBib}

\bibitem [\protect \citeauthoryear {%
Alwin%
}{%
Alwin%
}{%
{\protect \APACyear {1997}}%
}]{%
alwin_feeling_1997}
\APACinsertmetastar {%
alwin_feeling_1997}%
\begin{APACrefauthors}%
Alwin, D\BPBI F.%
\end{APACrefauthors}%
\unskip\
\newblock
\APACrefYearMonthDay{1997}{}{}.
\newblock
{\BBOQ}\APACrefatitle {Feeling {Thermometers} {Versus} 7-{Point} {Scales}: {Which} are {Better}?} {Feeling {Thermometers} {Versus} 7-{Point} {Scales}: {Which} are {Better}?}{\BBCQ}
\newblock
\APACjournalVolNumPages{Sociological Methods \& Research}{25}{3}{318--340}.
\newblock
\begin{APACrefDOI} \doi{10.1177/0049124197025003003} \end{APACrefDOI}
\PrintBackRefs{\CurrentBib}

\bibitem [\protect \citeauthoryear {%
Alwin%
\ \BBA {} Krosnick%
}{%
Alwin%
\ \BBA {} Krosnick%
}{%
{\protect \APACyear {1991}}%
}]{%
alwin_reliability_1991}
\APACinsertmetastar {%
alwin_reliability_1991}%
\begin{APACrefauthors}%
Alwin, D\BPBI F.%
\BCBT {}\ \BBA {} Krosnick, J\BPBI A.%
\end{APACrefauthors}%
\unskip\
\newblock
\APACrefYearMonthDay{1991}{}{}.
\newblock
{\BBOQ}\APACrefatitle {The {Reliability} of {Survey} {Attitude} {Measurement}: {The} {Influence} of {Question} and {Respondent} {Attributes}} {The {Reliability} of {Survey} {Attitude} {Measurement}: {The} {Influence} of {Question} and {Respondent} {Attributes}}.{\BBCQ}
\newblock
\APACjournalVolNumPages{Sociological Methods \& Research}{20}{1}{139--181}.
\newblock
\begin{APACrefDOI} \doi{10.1177/0049124191020001005} \end{APACrefDOI}
\PrintBackRefs{\CurrentBib}

\bibitem [\protect \citeauthoryear {%
Averbuch%
\ \BBA {} Katzper%
}{%
Averbuch%
\ \BBA {} Katzper%
}{%
{\protect \APACyear {2004}}%
}]{%
averbuch_assessment_2004}
\APACinsertmetastar {%
averbuch_assessment_2004}%
\begin{APACrefauthors}%
Averbuch, M.%
\BCBT {}\ \BBA {} Katzper, M.%
\end{APACrefauthors}%
\unskip\
\newblock
\APACrefYearMonthDay{2004}{}{}.
\newblock
{\BBOQ}\APACrefatitle {Assessment of {Visual} {Analog} versus {Categorical} {Scale} for {Measurement} of {Osteoarthritis} {Pain}} {Assessment of {Visual} {Analog} versus {Categorical} {Scale} for {Measurement} of {Osteoarthritis} {Pain}}.{\BBCQ}
\newblock
\APACjournalVolNumPages{The Journal of Clinical Pharmacology}{44}{4}{368--372}.
\newblock
\begin{APACrefDOI} \doi{10.1177/0091270004263995} \end{APACrefDOI}
\PrintBackRefs{\CurrentBib}

\bibitem [\protect \citeauthoryear {%
Baddeley%
}{%
Baddeley%
}{%
{\protect \APACyear {1994}}%
}]{%
baddeley_magical_1994}
\APACinsertmetastar {%
baddeley_magical_1994}%
\begin{APACrefauthors}%
Baddeley, A.%
\end{APACrefauthors}%
\unskip\
\newblock
\APACrefYearMonthDay{1994}{}{}.
\newblock
{\BBOQ}\APACrefatitle {The magical number seven: {Still} magic after all these years?} {The magical number seven: {Still} magic after all these years?}{\BBCQ}
\newblock
\APACjournalVolNumPages{Psychological Review}{101}{2}{535--356}.
\PrintBackRefs{\CurrentBib}

\bibitem [\protect \citeauthoryear {%
Bijur%
, Silver%
\BCBL {}\ \BBA {} Gallagher%
}{%
Bijur%
\ \protect \BOthers {.}}{%
{\protect \APACyear {2001}}%
}]{%
bijur_reliability_2001}
\APACinsertmetastar {%
bijur_reliability_2001}%
\begin{APACrefauthors}%
Bijur, P\BPBI E.%
, Silver, W.%
\BCBL {}\ \BBA {} Gallagher, E\BPBI J.%
\end{APACrefauthors}%
\unskip\
\newblock
\APACrefYearMonthDay{2001}{}{}.
\newblock
{\BBOQ}\APACrefatitle {Reliability of the {Visual} {Analog} {Scale} for {Measurement} of {Acute} {Pain}} {Reliability of the {Visual} {Analog} {Scale} for {Measurement} of {Acute} {Pain}}.{\BBCQ}
\newblock
\APACjournalVolNumPages{Academic Emergency Medicine}{8}{12}{1153--1157}.
\newblock
\begin{APACrefDOI} \doi{10.1111/j.1553-2712.2001.tb01132.x} \end{APACrefDOI}
\PrintBackRefs{\CurrentBib}

\bibitem [\protect \citeauthoryear {%
Bosch%
, Revilla%
, DeCastellarnau%
\BCBL {}\ \BBA {} Weber%
}{%
Bosch%
\ \protect \BOthers {.}}{%
{\protect \APACyear {2019}}%
}]{%
bosch_measurement_2019}
\APACinsertmetastar {%
bosch_measurement_2019}%
\begin{APACrefauthors}%
Bosch, O\BPBI J.%
, Revilla, M.%
, DeCastellarnau, A.%
\BCBL {}\ \BBA {} Weber, W.%
\end{APACrefauthors}%
\unskip\
\newblock
\APACrefYearMonthDay{2019}{}{}.
\newblock
{\BBOQ}\APACrefatitle {Measurement reliability, validity, and quality of slider versus radio button scales in an online probability-based panel in {Norway}} {Measurement reliability, validity, and quality of slider versus radio button scales in an online probability-based panel in {Norway}}.{\BBCQ}
\newblock
\APACjournalVolNumPages{Social Science Computer Review}{37}{1}{119--132}.
\PrintBackRefs{\CurrentBib}

\bibitem [\protect \citeauthoryear {%
Camarda%
, Eilers%
\BCBL {}\ \BBA {} Gampe%
}{%
Camarda%
\ \protect \BOthers {.}}{%
{\protect \APACyear {2008}}%
}]{%
camarda_modelling_2008}
\APACinsertmetastar {%
camarda_modelling_2008}%
\begin{APACrefauthors}%
Camarda, C\BPBI G.%
, Eilers, P\BPBI H.%
\BCBL {}\ \BBA {} Gampe, J.%
\end{APACrefauthors}%
\unskip\
\newblock
\APACrefYearMonthDay{2008}{}{}.
\newblock
{\BBOQ}\APACrefatitle {Modelling general patterns of digit preference} {Modelling general patterns of digit preference}.{\BBCQ}
\newblock
\APACjournalVolNumPages{Statistical Modelling}{8}{4}{385--401}.
\newblock
\begin{APACrefDOI} \doi{10.1177/1471082X0800800404} \end{APACrefDOI}
\PrintBackRefs{\CurrentBib}

\bibitem [\protect \citeauthoryear {%
Champney%
\ \BBA {} Marshall%
}{%
Champney%
\ \BBA {} Marshall%
}{%
{\protect \APACyear {1939}}%
}]{%
champney_optimal_1939}
\APACinsertmetastar {%
champney_optimal_1939}%
\begin{APACrefauthors}%
Champney, H.%
\BCBT {}\ \BBA {} Marshall, H.%
\end{APACrefauthors}%
\unskip\
\newblock
\APACrefYearMonthDay{1939}{}{}.
\newblock
{\BBOQ}\APACrefatitle {Optimal refinement of the rating scale.} {Optimal refinement of the rating scale.}{\BBCQ}
\newblock
\APACjournalVolNumPages{Journal of Applied Psychology}{23}{3}{323}.
\PrintBackRefs{\CurrentBib}

\bibitem [\protect \citeauthoryear {%
Davey%
, Barratt%
, Butow%
\BCBL {}\ \BBA {} Deeks%
}{%
Davey%
\ \protect \BOthers {.}}{%
{\protect \APACyear {2007}}%
}]{%
davey_one-item_2007}
\APACinsertmetastar {%
davey_one-item_2007}%
\begin{APACrefauthors}%
Davey, H\BPBI M.%
, Barratt, A\BPBI L.%
, Butow, P\BPBI N.%
\BCBL {}\ \BBA {} Deeks, J\BPBI J.%
\end{APACrefauthors}%
\unskip\
\newblock
\APACrefYearMonthDay{2007}{}{}.
\newblock
{\BBOQ}\APACrefatitle {A one-item question with a {Likert} or {Visual} {Analog} {Scale} adequately measured current anxiety} {A one-item question with a {Likert} or {Visual} {Analog} {Scale} adequately measured current anxiety}.{\BBCQ}
\newblock
\APACjournalVolNumPages{Journal of clinical epidemiology}{60}{4}{356--360}.
\PrintBackRefs{\CurrentBib}

\bibitem [\protect \citeauthoryear {%
Edward F.~Krieg%
}{%
Edward F.~Krieg%
}{%
{\protect \APACyear {1999}}%
}]{%
edward_f__krieg_biases_1999}
\APACinsertmetastar {%
edward_f__krieg_biases_1999}%
\begin{APACrefauthors}%
Edward F.~Krieg, J.%
\end{APACrefauthors}%
\unskip\
\newblock
\APACrefYearMonthDay{1999}{}{}.
\newblock
{\BBOQ}\APACrefatitle {Biases {Induced} by {Coarse} {Measurement} {Scales}} {Biases {Induced} by {Coarse} {Measurement} {Scales}}.{\BBCQ}
\newblock
\APACjournalVolNumPages{Educational and Psychological Measurement}{}{}{}.
\newblock
\begin{APACrefDOI} \doi{10.1177/00131649921970125} \end{APACrefDOI}
\PrintBackRefs{\CurrentBib}

\bibitem [\protect \citeauthoryear {%
Embretson%
\ \BBA {} Reise%
}{%
Embretson%
\ \BBA {} Reise%
}{%
{\protect \APACyear {2000}}%
}]{%
embretson_item_2000}
\APACinsertmetastar {%
embretson_item_2000}%
\begin{APACrefauthors}%
Embretson, S\BPBI E.%
\BCBT {}\ \BBA {} Reise, S\BPBI P.%
\end{APACrefauthors}%
\unskip\
\newblock
\APACrefYear{2000}.
\newblock
\APACrefbtitle {Item response theory for psychologists} {Item response theory for psychologists}.
\newblock
\APACaddressPublisher{Mahwah, NJ, US}{Lawrence Erlbaum Associates Publishers}.
\PrintBackRefs{\CurrentBib}

\bibitem [\protect \citeauthoryear {%
Ferrando%
}{%
Ferrando%
}{%
{\protect \APACyear {2003}}%
}]{%
ferrando_kernel_2003}
\APACinsertmetastar {%
ferrando_kernel_2003}%
\begin{APACrefauthors}%
Ferrando, P\BPBI J.%
\end{APACrefauthors}%
\unskip\
\newblock
\APACrefYearMonthDay{2003}{}{}.
\newblock
{\BBOQ}\APACrefatitle {A {Kernel} {Density} {Analysis} of {Continuous} {Typical}-{Response} {Scales}} {A {Kernel} {Density} {Analysis} of {Continuous} {Typical}-{Response} {Scales}}.{\BBCQ}
\newblock
\APACjournalVolNumPages{Educational and Psychological Measurement}{63}{5}{809--824}.
\newblock
\begin{APACrefDOI} \doi{10.1177/0013164403251323} \end{APACrefDOI}
\PrintBackRefs{\CurrentBib}

\bibitem [\protect \citeauthoryear {%
Flynn%
, Van~Schaik%
\BCBL {}\ \BBA {} Van~Wersch%
}{%
Flynn%
\ \protect \BOthers {.}}{%
{\protect \APACyear {2004}}%
}]{%
flynn_comparison_2004}
\APACinsertmetastar {%
flynn_comparison_2004}%
\begin{APACrefauthors}%
Flynn, D.%
, Van~Schaik, P.%
\BCBL {}\ \BBA {} Van~Wersch, A.%
\end{APACrefauthors}%
\unskip\
\newblock
\APACrefYearMonthDay{2004}{}{}.
\newblock
{\BBOQ}\APACrefatitle {A comparison of multi-item likert and visual analogue scales for the assessment of transactionally defined coping function1} {A comparison of multi-item likert and visual analogue scales for the assessment of transactionally defined coping function1}.{\BBCQ}
\newblock
\APACjournalVolNumPages{European Journal of Psychological Assessment}{20}{1}{49--58}.
\PrintBackRefs{\CurrentBib}

\bibitem [\protect \citeauthoryear {%
Funke%
\ \BBA {} Reips%
}{%
Funke%
\ \BBA {} Reips%
}{%
{\protect \APACyear {2012}}%
}]{%
funke_why_2012}
\APACinsertmetastar {%
funke_why_2012}%
\begin{APACrefauthors}%
Funke, F.%
\BCBT {}\ \BBA {} Reips, U\BHBI D.%
\end{APACrefauthors}%
\unskip\
\newblock
\APACrefYearMonthDay{2012}{}{}.
\newblock
{\BBOQ}\APACrefatitle {Why {Semantic} {Differentials} in {Web}-{Based} {Research} {Should} {Be} {Made} from {Visual} {Analogue} {Scales} and {Not} from 5-{Point} {Scales}} {Why {Semantic} {Differentials} in {Web}-{Based} {Research} {Should} {Be} {Made} from {Visual} {Analogue} {Scales} and {Not} from 5-{Point} {Scales}}.{\BBCQ}
\newblock
\APACjournalVolNumPages{Field Methods}{24}{3}{310--327}.
\newblock
\begin{APACrefDOI} \doi{10.1177/1525822X12444061} \end{APACrefDOI}
\PrintBackRefs{\CurrentBib}

\bibitem [\protect \citeauthoryear {%
Guyatt%
, Townsend%
, Berman%
\BCBL {}\ \BBA {} Keller%
}{%
Guyatt%
\ \protect \BOthers {.}}{%
{\protect \APACyear {1987}}%
}]{%
guyatt_comparison_1987}
\APACinsertmetastar {%
guyatt_comparison_1987}%
\begin{APACrefauthors}%
Guyatt, G\BPBI H.%
, Townsend, M.%
, Berman, L\BPBI B.%
\BCBL {}\ \BBA {} Keller, J\BPBI L.%
\end{APACrefauthors}%
\unskip\
\newblock
\APACrefYearMonthDay{1987}{}{}.
\newblock
{\BBOQ}\APACrefatitle {A comparison of {Likert} and visual analogue scales for measuring change in function} {A comparison of {Likert} and visual analogue scales for measuring change in function}.{\BBCQ}
\newblock
\APACjournalVolNumPages{Journal of chronic diseases}{40}{12}{1129--1133}.
\PrintBackRefs{\CurrentBib}

\bibitem [\protect \citeauthoryear {%
Haslbeck%
\ \protect \BOthers {.}}{%
Haslbeck%
\ \protect \BOthers {.}}{%
{\protect \APACyear {2024}}%
}]{%
haslbeck_comparing_2024}
\APACinsertmetastar {%
haslbeck_comparing_2024}%
\begin{APACrefauthors}%
Haslbeck, J.%
, Martínez, A\BPBI J.%
, Roefs, A.%
, Fried, E\BPBI I.%
, Lemmens, L\BPBI H\BPBI J\BPBI M.%
, Groot, E.%
\BCBL {}\ \BBA {} Edelsbrunner, P.%
\end{APACrefauthors}%
\unskip\
\newblock
\APACrefYearMonthDay{2024}{}{}.
\newblock
\APACrefbtitle {Comparing {Likert} and {Visual} {Analogue} {Scales} in {Ecological} {Momentary} {Assessment}.} {Comparing {Likert} and {Visual} {Analogue} {Scales} in {Ecological} {Momentary} {Assessment}.}
\newblock
\APACaddressPublisher{}{OSF}.
\newblock
\begin{APACrefDOI} \doi{10.31234/osf.io/yt8xw} \end{APACrefDOI}
\PrintBackRefs{\CurrentBib}

\bibitem [\protect \citeauthoryear {%
Hasson%
\ \BBA {} Arnetz%
}{%
Hasson%
\ \BBA {} Arnetz%
}{%
{\protect \APACyear {2005}}%
}]{%
hasson_validation_2005}
\APACinsertmetastar {%
hasson_validation_2005}%
\begin{APACrefauthors}%
Hasson, D.%
\BCBT {}\ \BBA {} Arnetz, B\BPBI B.%
\end{APACrefauthors}%
\unskip\
\newblock
\APACrefYearMonthDay{2005}{}{}.
\newblock
{\BBOQ}\APACrefatitle {Validation and findings comparing {VAS} vs. {Likert} scales for psychosocial measurements.} {Validation and findings comparing {VAS} vs. {Likert} scales for psychosocial measurements.}{\BBCQ}
\newblock
\APACjournalVolNumPages{International Electronic Journal of Health Education}{8}{}{178--192}.
\PrintBackRefs{\CurrentBib}

\bibitem [\protect \citeauthoryear {%
Hayes%
}{%
Hayes%
}{%
{\protect \APACyear {1921}}%
}]{%
hayes_experimental_1921}
\APACinsertmetastar {%
hayes_experimental_1921}%
\begin{APACrefauthors}%
Hayes, M\BPBI H.%
\end{APACrefauthors}%
\unskip\
\newblock
\APACrefYearMonthDay{1921}{}{}.
\newblock
{\BBOQ}\APACrefatitle {Experimental development of the graphic rating method} {Experimental development of the graphic rating method}.{\BBCQ}
\newblock
\APACjournalVolNumPages{Psychological Bulletin}{18}{}{98--99}.
\PrintBackRefs{\CurrentBib}

\bibitem [\protect \citeauthoryear {%
Hilbert%
}{%
Hilbert%
}{%
{\protect \APACyear {2015}}%
}]{%
hilbert_influence_2015}
\APACinsertmetastar {%
hilbert_influence_2015}%
\begin{APACrefauthors}%
Hilbert, S.%
\end{APACrefauthors}%
\unskip\
\newblock
\APACrefYearMonthDay{2015}{}{}.
\newblock
{\BBOQ}\APACrefatitle {The influence of the response format in a personality questionnaire: {An} analysis of a dichotomous, a {Likert}-type, and a visual analogue scale} {The influence of the response format in a personality questionnaire: {An} analysis of a dichotomous, a {Likert}-type, and a visual analogue scale}.{\BBCQ}
\newblock
\APACjournalVolNumPages{TPM - Testing, Psychometrics, Methodology in Applied Psychology}{}{1}{3--24}.
\newblock
\begin{APACrefDOI} \doi{10.4473/TPM23.1.1} \end{APACrefDOI}
\PrintBackRefs{\CurrentBib}

\bibitem [\protect \citeauthoryear {%
Jacob%
, Thomas%
\BCBL {}\ \BBA {} Joseph%
}{%
Jacob%
\ \protect \BOthers {.}}{%
{\protect \APACyear {2024}}%
}]{%
jacob_over_2024}
\APACinsertmetastar {%
jacob_over_2024}%
\begin{APACrefauthors}%
Jacob, B\BPBI M.%
, Thomas, S.%
\BCBL {}\ \BBA {} Joseph, J.%
\end{APACrefauthors}%
\unskip\
\newblock
\APACrefYearMonthDay{2024}{}{}.
\newblock
{\BBOQ}\APACrefatitle {Over two decades of research on choice overload: {An} overview and research agenda} {Over two decades of research on choice overload: {An} overview and research agenda}.{\BBCQ}
\newblock
\APACjournalVolNumPages{International Journal of Consumer Studies}{48}{2}{e13029}.
\newblock
\begin{APACrefDOI} \doi{10.1111/ijcs.13029} \end{APACrefDOI}
\PrintBackRefs{\CurrentBib}

\bibitem [\protect \citeauthoryear {%
Jaeschke%
, Singer%
\BCBL {}\ \BBA {} Guyatt%
}{%
Jaeschke%
\ \protect \BOthers {.}}{%
{\protect \APACyear {1990}}%
}]{%
jaeschke_comparison_1990}
\APACinsertmetastar {%
jaeschke_comparison_1990}%
\begin{APACrefauthors}%
Jaeschke, R.%
, Singer, J.%
\BCBL {}\ \BBA {} Guyatt, G\BPBI H.%
\end{APACrefauthors}%
\unskip\
\newblock
\APACrefYearMonthDay{1990}{}{}.
\newblock
{\BBOQ}\APACrefatitle {A comparison of seven-point and visual analogue scales: data from a randomized trial} {A comparison of seven-point and visual analogue scales: data from a randomized trial}.{\BBCQ}
\newblock
\APACjournalVolNumPages{Controlled clinical trials}{11}{1}{43--51}.
\PrintBackRefs{\CurrentBib}

\bibitem [\protect \citeauthoryear {%
John%
\ \BBA {} Srivastava%
}{%
John%
\ \BBA {} Srivastava%
}{%
{\protect \APACyear {1999}}%
}]{%
john_big-five_1999}
\APACinsertmetastar {%
john_big-five_1999}%
\begin{APACrefauthors}%
John, O\BPBI P.%
\BCBT {}\ \BBA {} Srivastava, S.%
\end{APACrefauthors}%
\unskip\
\newblock
\APACrefYearMonthDay{1999}{}{}.
\newblock
{\BBOQ}\APACrefatitle {The {Big} {Five} {Trait} taxonomy: {History}, measurement, and theoretical perspectives} {The {Big} {Five} {Trait} taxonomy: {History}, measurement, and theoretical perspectives}.{\BBCQ}
\newblock
\BIn{} \APACrefbtitle {Handbook of personality: {Theory} and research, 2nd ed} {Handbook of personality: {Theory} and research, 2nd ed}\ (\BPGS\ 102--138).
\newblock
\APACaddressPublisher{New York, NY, US}{Guilford Press}.
\PrintBackRefs{\CurrentBib}

\bibitem [\protect \citeauthoryear {%
Kuhlmann%
, Dantlgraber%
\BCBL {}\ \BBA {} Reips%
}{%
Kuhlmann%
\ \protect \BOthers {.}}{%
{\protect \APACyear {2017}}%
}]{%
kuhlmann_investigating_2017}
\APACinsertmetastar {%
kuhlmann_investigating_2017}%
\begin{APACrefauthors}%
Kuhlmann, T.%
, Dantlgraber, M.%
\BCBL {}\ \BBA {} Reips, U\BHBI D.%
\end{APACrefauthors}%
\unskip\
\newblock
\APACrefYearMonthDay{2017}{}{}.
\newblock
{\BBOQ}\APACrefatitle {Investigating measurement equivalence of visual analogue scales and {Likert}-type scales in {Internet}-based personality questionnaires} {Investigating measurement equivalence of visual analogue scales and {Likert}-type scales in {Internet}-based personality questionnaires}.{\BBCQ}
\newblock
\APACjournalVolNumPages{Behavior Research Methods}{49}{6}{2173--2181}.
\newblock
\begin{APACrefDOI} \doi{10.3758/s13428-016-0850-x} \end{APACrefDOI}
\PrintBackRefs{\CurrentBib}

\bibitem [\protect \citeauthoryear {%
Lee%
\ \BBA {} Paek%
}{%
Lee%
\ \BBA {} Paek%
}{%
{\protect \APACyear {2014}}%
}]{%
lee_search_2014}
\APACinsertmetastar {%
lee_search_2014}%
\begin{APACrefauthors}%
Lee, J.%
\BCBT {}\ \BBA {} Paek, I.%
\end{APACrefauthors}%
\unskip\
\newblock
\APACrefYearMonthDay{2014}{}{}.
\newblock
{\BBOQ}\APACrefatitle {In {Search} of the {Optimal} {Number} of {Response} {Categories} in a {Rating} {Scale}} {In {Search} of the {Optimal} {Number} of {Response} {Categories} in a {Rating} {Scale}}.{\BBCQ}
\newblock
\APACjournalVolNumPages{Journal of Psychoeducational Assessment}{32}{7}{663--673}.
\newblock
\begin{APACrefDOI} \doi{10.1177/0734282914522200} \end{APACrefDOI}
\PrintBackRefs{\CurrentBib}

\bibitem [\protect \citeauthoryear {%
{Lei Chang}%
}{%
{Lei Chang}%
}{%
{\protect \APACyear {1994}}%
}]{%
lei_chang_psychometric_1994}
\APACinsertmetastar {%
lei_chang_psychometric_1994}%
\begin{APACrefauthors}%
{Lei Chang}.%
\end{APACrefauthors}%
\unskip\
\newblock
\APACrefYearMonthDay{1994}{}{}.
\newblock
{\BBOQ}\APACrefatitle {A {Psychometric} {Evaluation} of 4-{Point} and 6-{Point} {Likert}-{Type} {Scales} in {Relation} to {Reliability} and {Validity}} {A {Psychometric} {Evaluation} of 4-{Point} and 6-{Point} {Likert}-{Type} {Scales} in {Relation} to {Reliability} and {Validity}}.{\BBCQ}
\newblock
\APACjournalVolNumPages{Applied Psychological Measurement}{18}{3}{205--215}.
\newblock
\begin{APACrefDOI} \doi{10.1177/014662169401800302} \end{APACrefDOI}
\PrintBackRefs{\CurrentBib}

\bibitem [\protect \citeauthoryear {%
Lesage%
, Berjot%
\BCBL {}\ \BBA {} Deschamps%
}{%
Lesage%
\ \protect \BOthers {.}}{%
{\protect \APACyear {2012}}%
}]{%
lesage_clinical_2012}
\APACinsertmetastar {%
lesage_clinical_2012}%
\begin{APACrefauthors}%
Lesage, F\BHBI X.%
, Berjot, S.%
\BCBL {}\ \BBA {} Deschamps, F.%
\end{APACrefauthors}%
\unskip\
\newblock
\APACrefYearMonthDay{2012}{}{}.
\newblock
{\BBOQ}\APACrefatitle {Clinical stress assessment using a visual analogue scale} {Clinical stress assessment using a visual analogue scale}.{\BBCQ}
\newblock
\APACjournalVolNumPages{Occupational medicine}{62}{8}{600--605}.
\PrintBackRefs{\CurrentBib}

\bibitem [\protect \citeauthoryear {%
Likert%
}{%
Likert%
}{%
{\protect \APACyear {1932}}%
}]{%
likert_technique_1932}
\APACinsertmetastar {%
likert_technique_1932}%
\begin{APACrefauthors}%
Likert, R.%
\end{APACrefauthors}%
\unskip\
\newblock
\APACrefYearMonthDay{1932}{}{}.
\newblock
{\BBOQ}\APACrefatitle {A technique for the measurement of attitudes} {A technique for the measurement of attitudes}.{\BBCQ}
\newblock
\APACjournalVolNumPages{Archives of Psychology}{}{}{}.
\PrintBackRefs{\CurrentBib}

\bibitem [\protect \citeauthoryear {%
Loevinger%
}{%
Loevinger%
}{%
{\protect \APACyear {1954}}%
}]{%
loevinger_attenuation_1954}
\APACinsertmetastar {%
loevinger_attenuation_1954}%
\begin{APACrefauthors}%
Loevinger, J.%
\end{APACrefauthors}%
\unskip\
\newblock
\APACrefYearMonthDay{1954}{}{}.
\newblock
{\BBOQ}\APACrefatitle {The attenuation paradox in test theory} {The attenuation paradox in test theory}.{\BBCQ}
\newblock
\APACjournalVolNumPages{Psychological Bulletin}{51}{5}{493--504}.
\newblock
\begin{APACrefDOI} \doi{10.1037/h0058543} \end{APACrefDOI}
\PrintBackRefs{\CurrentBib}

\bibitem [\protect \citeauthoryear {%
Lord%
\ \BBA {} Novick%
}{%
Lord%
\ \BBA {} Novick%
}{%
{\protect \APACyear {1968}}%
}]{%
lord_statistical_1968}
\APACinsertmetastar {%
lord_statistical_1968}%
\begin{APACrefauthors}%
Lord, F\BPBI M.%
\BCBT {}\ \BBA {} Novick, M\BPBI R.%
\end{APACrefauthors}%
\unskip\
\newblock
\APACrefYear{1968}.
\newblock
\APACrefbtitle {Statistical theories of mental test scores} {Statistical theories of mental test scores}.
\newblock
\APACaddressPublisher{}{IAP}.
\PrintBackRefs{\CurrentBib}

\bibitem [\protect \citeauthoryear {%
Lozano%
, García-Cueto%
\BCBL {}\ \BBA {} Muñiz%
}{%
Lozano%
\ \protect \BOthers {.}}{%
{\protect \APACyear {2008}}%
}]{%
lozano_effect_2008}
\APACinsertmetastar {%
lozano_effect_2008}%
\begin{APACrefauthors}%
Lozano, L\BPBI M.%
, García-Cueto, E.%
\BCBL {}\ \BBA {} Muñiz, J.%
\end{APACrefauthors}%
\unskip\
\newblock
\APACrefYearMonthDay{2008}{}{}.
\newblock
{\BBOQ}\APACrefatitle {Effect of the {Number} of {Response} {Categories} on the {Reliability} and {Validity} of {Rating} {Scales}} {Effect of the {Number} of {Response} {Categories} on the {Reliability} and {Validity} of {Rating} {Scales}}.{\BBCQ}
\newblock
\APACjournalVolNumPages{Methodology}{4}{2}{73--79}.
\newblock
\begin{APACrefDOI} \doi{10.1027/1614-2241.4.2.73} \end{APACrefDOI}
\PrintBackRefs{\CurrentBib}

\bibitem [\protect \citeauthoryear {%
Mart\'inez%
, Lemmens%
, Fried%
, Gu\"omundsdóttir%
\BCBL {}\ \BBA {} Roefs%
}{%
Mart\'inez%
\ \protect \BOthers {.}}{%
{\protect \APACyear {2024}}%
}]{%
martinez_validation_2024}
\APACinsertmetastar {%
martinez_validation_2024}%
\begin{APACrefauthors}%
Mart\'inez, A\BPBI J.%
, Lemmens, L\BPBI H\BPBI J\BPBI M.%
, Fried, E\BPBI I.%
, Gu\"omundsdóttir, G\BPBI R.%
\BCBL {}\ \BBA {} Roefs, A.%
\end{APACrefauthors}%
\unskip\
\newblock
\APACrefYearMonthDay{2024}{}{}.
\newblock
\APACrefbtitle {Validation of a transdiagnostic psychopathology {EMA} protocol in a university students sample.} {Validation of a transdiagnostic psychopathology {EMA} protocol in a university students sample.}
\newblock
\APACaddressPublisher{}{OSF}.
\newblock
\begin{APACrefDOI} \doi{10.31234/osf.io/5a6uz} \end{APACrefDOI}
\PrintBackRefs{\CurrentBib}

\bibitem [\protect \citeauthoryear {%
Martínez%
, Lemmens%
, Fried%
\BCBL {}\ \BBA {} Roefs%
}{%
Martínez%
\ \protect \BOthers {.}}{%
{\protect \APACyear {2023}}%
}]{%
martinez_developing_2023}
\APACinsertmetastar {%
martinez_developing_2023}%
\begin{APACrefauthors}%
Martínez, A\BPBI J.%
, Lemmens, L.%
, Fried, E\BPBI I.%
\BCBL {}\ \BBA {} Roefs, A.%
\end{APACrefauthors}%
\unskip\
\newblock
\APACrefYearMonthDay{2023}{}{}.
\newblock
\APACrefbtitle {Developing a {Transdiagnostic} {Ecological} {Momentary} {Assessment} {Protocol} for {Psychopathology}.} {Developing a {Transdiagnostic} {Ecological} {Momentary} {Assessment} {Protocol} for {Psychopathology}.}
\newblock
\APACaddressPublisher{}{OSF}.
\newblock
\begin{APACrefDOI} \doi{10.31234/osf.io/nvf89} \end{APACrefDOI}
\PrintBackRefs{\CurrentBib}

\bibitem [\protect \citeauthoryear {%
Maydeu-Olivares%
, Kramp%
, García-Forero%
, Gallardo-Pujol%
\BCBL {}\ \BBA {} Coffman%
}{%
Maydeu-Olivares%
\ \protect \BOthers {.}}{%
{\protect \APACyear {2009}}%
}]{%
maydeu-olivares_effect_2009}
\APACinsertmetastar {%
maydeu-olivares_effect_2009}%
\begin{APACrefauthors}%
Maydeu-Olivares, A.%
, Kramp, U.%
, García-Forero, C.%
, Gallardo-Pujol, D.%
\BCBL {}\ \BBA {} Coffman, D.%
\end{APACrefauthors}%
\unskip\
\newblock
\APACrefYearMonthDay{2009}{}{}.
\newblock
{\BBOQ}\APACrefatitle {The effect of varying the number of response alternatives in rating scales: {Experimental} evidence from intra-individual effects} {The effect of varying the number of response alternatives in rating scales: {Experimental} evidence from intra-individual effects}.{\BBCQ}
\newblock
\APACjournalVolNumPages{Behavior Research Methods}{41}{2}{295--308}.
\newblock
\begin{APACrefDOI} \doi{10.3758/BRM.41.2.295} \end{APACrefDOI}
\PrintBackRefs{\CurrentBib}

\bibitem [\protect \citeauthoryear {%
McArdle%
}{%
McArdle%
}{%
{\protect \APACyear {2009}}%
}]{%
mcardle_latent_2009}
\APACinsertmetastar {%
mcardle_latent_2009}%
\begin{APACrefauthors}%
McArdle, J\BPBI J.%
\end{APACrefauthors}%
\unskip\
\newblock
\APACrefYearMonthDay{2009}{}{}.
\newblock
{\BBOQ}\APACrefatitle {Latent {Variable} {Modeling} of {Differences} and {Changes} with {Longitudinal} {Data}} {Latent {Variable} {Modeling} of {Differences} and {Changes} with {Longitudinal} {Data}}.{\BBCQ}
\newblock
\APACjournalVolNumPages{Annual Review of Psychology}{60}{1}{577--605}.
\newblock
\begin{APACrefDOI} \doi{10.1146/annurev.psych.60.110707.163612} \end{APACrefDOI}
\PrintBackRefs{\CurrentBib}

\bibitem [\protect \citeauthoryear {%
McGrath%
, Mitchell%
, Kim%
\BCBL {}\ \BBA {} Hough%
}{%
McGrath%
\ \protect \BOthers {.}}{%
{\protect \APACyear {2010}}%
}]{%
mcgrath_evidence_2010}
\APACinsertmetastar {%
mcgrath_evidence_2010}%
\begin{APACrefauthors}%
McGrath, R\BPBI E.%
, Mitchell, M.%
, Kim, B\BPBI H.%
\BCBL {}\ \BBA {} Hough, L.%
\end{APACrefauthors}%
\unskip\
\newblock
\APACrefYearMonthDay{2010}{}{}.
\newblock
{\BBOQ}\APACrefatitle {Evidence for response bias as a source of error variance in applied assessment.} {Evidence for response bias as a source of error variance in applied assessment.}{\BBCQ}
\newblock
\APACjournalVolNumPages{Psychological bulletin}{136}{3}{450}.
\PrintBackRefs{\CurrentBib}

\bibitem [\protect \citeauthoryear {%
McKelvie%
}{%
McKelvie%
}{%
{\protect \APACyear {1978}}%
}]{%
mckelvie_graphic_1978}
\APACinsertmetastar {%
mckelvie_graphic_1978}%
\begin{APACrefauthors}%
McKelvie, S\BPBI J.%
\end{APACrefauthors}%
\unskip\
\newblock
\APACrefYearMonthDay{1978}{}{}.
\newblock
{\BBOQ}\APACrefatitle {Graphic rating scales — {How} many categories?} {Graphic rating scales — {How} many categories?}{\BBCQ}
\newblock
\APACjournalVolNumPages{British Journal of Psychology}{69}{2}{185--202}.
\newblock
\begin{APACrefDOI} \doi{10.1111/j.2044-8295.1978.tb01647.x} \end{APACrefDOI}
\PrintBackRefs{\CurrentBib}

\bibitem [\protect \citeauthoryear {%
Miller%
}{%
Miller%
}{%
{\protect \APACyear {1956}}%
}]{%
miller_magical_1956}
\APACinsertmetastar {%
miller_magical_1956}%
\begin{APACrefauthors}%
Miller, G\BPBI A.%
\end{APACrefauthors}%
\unskip\
\newblock
\APACrefYearMonthDay{1956}{}{}.
\newblock
{\BBOQ}\APACrefatitle {The magical number seven, plus or minus two: {Some} limits on our capacity for processing information} {The magical number seven, plus or minus two: {Some} limits on our capacity for processing information}.{\BBCQ}
\newblock
\APACjournalVolNumPages{Psychological Review}{63}{2}{81--97}.
\newblock
\begin{APACrefDOI} \doi{10.1037/h0043158} \end{APACrefDOI}
\PrintBackRefs{\CurrentBib}

\bibitem [\protect \citeauthoryear {%
Mu\~niz%
, Garc\'ia-Cueto%
\BCBL {}\ \BBA {} Lozano%
}{%
Mu\~niz%
\ \protect \BOthers {.}}{%
{\protect \APACyear {2005}}%
}]{%
muniz_item_2005}
\APACinsertmetastar {%
muniz_item_2005}%
\begin{APACrefauthors}%
Mu\~niz, J.%
, Garc\'ia-Cueto, E.%
\BCBL {}\ \BBA {} Lozano, L\BPBI M.%
\end{APACrefauthors}%
\unskip\
\newblock
\APACrefYearMonthDay{2005}{}{}.
\newblock
{\BBOQ}\APACrefatitle {Item format and the psychometric properties of the {Eysenck} {Personality} {Questionnaire}} {Item format and the psychometric properties of the {Eysenck} {Personality} {Questionnaire}}.{\BBCQ}
\newblock
\APACjournalVolNumPages{Personality and Individual Differences}{38}{1}{61--69}.
\newblock
\begin{APACrefDOI} \doi{10.1016/j.paid.2004.03.021} \end{APACrefDOI}
\PrintBackRefs{\CurrentBib}

\bibitem [\protect \citeauthoryear {%
Olsson%
, Drasgow%
\BCBL {}\ \BBA {} Dorans%
}{%
Olsson%
\ \protect \BOthers {.}}{%
{\protect \APACyear {1982}}%
}]{%
olsson_polyserial_1982}
\APACinsertmetastar {%
olsson_polyserial_1982}%
\begin{APACrefauthors}%
Olsson, U.%
, Drasgow, F.%
\BCBL {}\ \BBA {} Dorans, N\BPBI J.%
\end{APACrefauthors}%
\unskip\
\newblock
\APACrefYearMonthDay{1982}{}{}.
\newblock
{\BBOQ}\APACrefatitle {The polyserial correlation coefficient} {The polyserial correlation coefficient}.{\BBCQ}
\newblock
\APACjournalVolNumPages{Psychometrika}{47}{3}{337--347}.
\newblock
\begin{APACrefDOI} \doi{10.1007/BF02294164} \end{APACrefDOI}
\PrintBackRefs{\CurrentBib}

\bibitem [\protect \citeauthoryear {%
Paulhus%
}{%
Paulhus%
}{%
{\protect \APACyear {1991}}%
}]{%
paulhus_measurement_1991}
\APACinsertmetastar {%
paulhus_measurement_1991}%
\begin{APACrefauthors}%
Paulhus, D\BPBI L.%
\end{APACrefauthors}%
\unskip\
\newblock
\APACrefYearMonthDay{1991}{}{}.
\newblock
{\BBOQ}\APACrefatitle {Measurement and {Control} of {Response} {Bias}} {Measurement and {Control} of {Response} {Bias}}.{\BBCQ}
\newblock
\BIn{} J\BPBI P.~Robinson, P\BPBI R.~Shaver\BCBL {}\ \BBA {} L\BPBI S.~Wrightsman\ (\BEDS), \APACrefbtitle {Measures of {Personality} and {Social} {Psychological} {Attitudes}} {Measures of {Personality} and {Social} {Psychological} {Attitudes}}\ (\BPGS\ 17--59).
\newblock
\APACaddressPublisher{}{Academic Press}.
\newblock
\begin{APACrefDOI} \doi{10.1016/B978-0-12-590241-0.50006-X} \end{APACrefDOI}
\PrintBackRefs{\CurrentBib}

\bibitem [\protect \citeauthoryear {%
Podsakoff%
, MacKenzie%
, Lee%
\BCBL {}\ \BBA {} Podsakoff%
}{%
Podsakoff%
\ \protect \BOthers {.}}{%
{\protect \APACyear {2003}}%
}]{%
podsakoff_common_2003}
\APACinsertmetastar {%
podsakoff_common_2003}%
\begin{APACrefauthors}%
Podsakoff, P\BPBI M.%
, MacKenzie, S\BPBI B.%
, Lee, J\BHBI Y.%
\BCBL {}\ \BBA {} Podsakoff, N\BPBI P.%
\end{APACrefauthors}%
\unskip\
\newblock
\APACrefYearMonthDay{2003}{}{}.
\newblock
{\BBOQ}\APACrefatitle {Common method biases in behavioral research: a critical review of the literature and recommended remedies.} {Common method biases in behavioral research: a critical review of the literature and recommended remedies.}{\BBCQ}
\newblock
\APACjournalVolNumPages{Journal of applied psychology}{88}{5}{879}.
\PrintBackRefs{\CurrentBib}

\bibitem [\protect \citeauthoryear {%
Preston%
\ \BBA {} Colman%
}{%
Preston%
\ \BBA {} Colman%
}{%
{\protect \APACyear {2000}}%
}]{%
preston_optimal_2000}
\APACinsertmetastar {%
preston_optimal_2000}%
\begin{APACrefauthors}%
Preston, C\BPBI C.%
\BCBT {}\ \BBA {} Colman, A\BPBI M.%
\end{APACrefauthors}%
\unskip\
\newblock
\APACrefYearMonthDay{2000}{}{}.
\newblock
{\BBOQ}\APACrefatitle {Optimal number of response categories in rating scales: reliability, validity, discriminating power, and respondent preferences} {Optimal number of response categories in rating scales: reliability, validity, discriminating power, and respondent preferences}.{\BBCQ}
\newblock
\APACjournalVolNumPages{Acta Psychologica}{104}{1}{1--15}.
\newblock
\begin{APACrefDOI} \doi{10.1016/S0001-6918(99)00050-5} \end{APACrefDOI}
\PrintBackRefs{\CurrentBib}

\bibitem [\protect \citeauthoryear {%
Samejima%
}{%
Samejima%
}{%
{\protect \APACyear {1969}}%
}]{%
samejima_estimation_1969}
\APACinsertmetastar {%
samejima_estimation_1969}%
\begin{APACrefauthors}%
Samejima, F.%
\end{APACrefauthors}%
\unskip\
\newblock
\APACrefYear{1969}.
\newblock
\APACrefbtitle {Estimation of latent ability using a response pattern of graded scores} {Estimation of latent ability using a response pattern of graded scores}.
\newblock
\APACaddressPublisher{}{Psychometrika Society}.
\PrintBackRefs{\CurrentBib}

\bibitem [\protect \citeauthoryear {%
Scherpenzeel%
\ \BBA {} Saris%
}{%
Scherpenzeel%
\ \BBA {} Saris%
}{%
{\protect \APACyear {1997}}%
}]{%
scherpenzeel_validity_1997}
\APACinsertmetastar {%
scherpenzeel_validity_1997}%
\begin{APACrefauthors}%
Scherpenzeel, A\BPBI C.%
\BCBT {}\ \BBA {} Saris, W\BPBI E.%
\end{APACrefauthors}%
\unskip\
\newblock
\APACrefYearMonthDay{1997}{}{}.
\newblock
{\BBOQ}\APACrefatitle {The {Validity} and {Reliability} of {Survey} {Questions}: {A} {Meta}-{Analysis} of {MTMM} {Studies}} {The {Validity} and {Reliability} of {Survey} {Questions}: {A} {Meta}-{Analysis} of {MTMM} {Studies}}.{\BBCQ}
\newblock
\APACjournalVolNumPages{Sociological Methods \& Research}{25}{3}{341--383}.
\newblock
\begin{APACrefDOI} \doi{10.1177/0049124197025003004} \end{APACrefDOI}
\PrintBackRefs{\CurrentBib}

\bibitem [\protect \citeauthoryear {%
Schmidt%
}{%
Schmidt%
}{%
{\protect \APACyear {2010}}%
}]{%
schmidt_more_2010}
\APACinsertmetastar {%
schmidt_more_2010}%
\begin{APACrefauthors}%
Schmidt, K\BPBI M.%
\end{APACrefauthors}%
\unskip\
\newblock
\APACrefYearMonthDay{2010}{}{}.
\newblock
{\BBOQ}\APACrefatitle {More is {Not} {Better}: {Rescoring} {Combinations} for {Length} {Rating} {Scales}} {More is {Not} {Better}: {Rescoring} {Combinations} for {Length} {Rating} {Scales}}.{\BBCQ}
\newblock
\BIn{} \APACrefbtitle {International {Conference} on {Measurement} ({ICOM}-2010).} {International {Conference} on {Measurement} ({ICOM}-2010).}
\newblock
\APACaddressPublisher{Bethesda, MD}{}.
\PrintBackRefs{\CurrentBib}

\bibitem [\protect \citeauthoryear {%
Simms%
, Zelazny%
, Williams%
\BCBL {}\ \BBA {} Bernstein%
}{%
Simms%
\ \protect \BOthers {.}}{%
{\protect \APACyear {2019}}%
}]{%
simms_does_2019}
\APACinsertmetastar {%
simms_does_2019}%
\begin{APACrefauthors}%
Simms, L\BPBI J.%
, Zelazny, K.%
, Williams, T\BPBI F.%
\BCBL {}\ \BBA {} Bernstein, L.%
\end{APACrefauthors}%
\unskip\
\newblock
\APACrefYearMonthDay{2019}{}{}.
\newblock
{\BBOQ}\APACrefatitle {Does the number of response options matter? {Psychometric} perspectives using personality questionnaire data.} {Does the number of response options matter? {Psychometric} perspectives using personality questionnaire data.}{\BBCQ}
\newblock
\APACjournalVolNumPages{Psychological assessment}{31}{4}{557}.
\PrintBackRefs{\CurrentBib}

\bibitem [\protect \citeauthoryear {%
Sung%
\ \BBA {} Wu%
}{%
Sung%
\ \BBA {} Wu%
}{%
{\protect \APACyear {2018}}%
}]{%
sung_visual_2018}
\APACinsertmetastar {%
sung_visual_2018}%
\begin{APACrefauthors}%
Sung, Y\BHBI T.%
\BCBT {}\ \BBA {} Wu, J\BHBI S.%
\end{APACrefauthors}%
\unskip\
\newblock
\APACrefYearMonthDay{2018}{}{}.
\newblock
{\BBOQ}\APACrefatitle {The {Visual} {Analogue} {Scale} for {Rating}, {Ranking} and {Paired}-{Comparison} ({VAS}-{RRP}): {A} new technique for psychological measurement} {The {Visual} {Analogue} {Scale} for {Rating}, {Ranking} and {Paired}-{Comparison} ({VAS}-{RRP}): {A} new technique for psychological measurement}.{\BBCQ}
\newblock
\APACjournalVolNumPages{Behavior Research Methods}{50}{4}{1694--1715}.
\newblock
\begin{APACrefDOI} \doi{10.3758/s13428-018-1041-8} \end{APACrefDOI}
\PrintBackRefs{\CurrentBib}

\bibitem [\protect \citeauthoryear {%
Thomas%
, Uldall%
\BCBL {}\ \BBA {} Krosnick%
}{%
Thomas%
\ \protect \BOthers {.}}{%
{\protect \APACyear {2004}}%
}]{%
thomas_how_2004}
\APACinsertmetastar {%
thomas_how_2004}%
\begin{APACrefauthors}%
Thomas, R\BPBI K.%
, Uldall, B.%
\BCBL {}\ \BBA {} Krosnick, J\BPBI A.%
\end{APACrefauthors}%
\unskip\
\newblock
\APACrefYearMonthDay{2004}{}{}.
\newblock
{\BBOQ}\APACrefatitle {How many are too many? {Number} of response categories and validity} {How many are too many? {Number} of response categories and validity}.{\BBCQ}
\newblock
\BIn{} \APACrefbtitle {59th {Annual} {Conference} of the {American} {Association} of {Public} {Opinion} {Research}.} {59th {Annual} {Conference} of the {American} {Association} of {Public} {Opinion} {Research}.}
\PrintBackRefs{\CurrentBib}

\bibitem [\protect \citeauthoryear {%
Van~Laerhoven%
, Van Der~Zaag‐Loonen%
\BCBL {}\ \BBA {} Derkx%
}{%
Van~Laerhoven%
\ \protect \BOthers {.}}{%
{\protect \APACyear {2004}}%
}]{%
van_laerhoven_comparison_2004}
\APACinsertmetastar {%
van_laerhoven_comparison_2004}%
\begin{APACrefauthors}%
Van~Laerhoven, H.%
, Van Der~Zaag‐Loonen, H.%
\BCBL {}\ \BBA {} Derkx, B.%
\end{APACrefauthors}%
\unskip\
\newblock
\APACrefYearMonthDay{2004}{}{}.
\newblock
{\BBOQ}\APACrefatitle {A comparison of {Likert} scale and visual analogue scales as response options in children's questionnaires} {A comparison of {Likert} scale and visual analogue scales as response options in children's questionnaires}.{\BBCQ}
\newblock
\APACjournalVolNumPages{Acta Paediatrica}{93}{6}{830--835}.
\newblock
\begin{APACrefDOI} \doi{10.1111/j.1651-2227.2004.tb03026.x} \end{APACrefDOI}
\PrintBackRefs{\CurrentBib}

\bibitem [\protect \citeauthoryear {%
Viswanathan%
, Bergen%
, Dutta%
\BCBL {}\ \BBA {} Childers%
}{%
Viswanathan%
\ \protect \BOthers {.}}{%
{\protect \APACyear {1996}}%
}]{%
viswanathan_does_1996}
\APACinsertmetastar {%
viswanathan_does_1996}%
\begin{APACrefauthors}%
Viswanathan, M.%
, Bergen, M.%
, Dutta, S.%
\BCBL {}\ \BBA {} Childers, T.%
\end{APACrefauthors}%
\unskip\
\newblock
\APACrefYearMonthDay{1996}{}{}.
\newblock
{\BBOQ}\APACrefatitle {Does a single response category in a scale completely capture a response?} {Does a single response category in a scale completely capture a response?}{\BBCQ}
\newblock
\APACjournalVolNumPages{Psychology \& Marketing}{13}{5}{457--479}.
\newblock
\begin{APACrefDOI} \doi{10.1002/(SICI)1520-6793(199608)13:5<457::AID-MAR2>3.0.CO;2-8} \end{APACrefDOI}
\PrintBackRefs{\CurrentBib}

\bibitem [\protect \citeauthoryear {%
Wang%
, Zhang%
, McArdle%
\BCBL {}\ \BBA {} Salthouse%
}{%
Wang%
\ \protect \BOthers {.}}{%
{\protect \APACyear {2008}}%
}]{%
wang_investigating_2008}
\APACinsertmetastar {%
wang_investigating_2008}%
\begin{APACrefauthors}%
Wang, L.%
, Zhang, Z.%
, McArdle, J\BPBI J.%
\BCBL {}\ \BBA {} Salthouse, T\BPBI A.%
\end{APACrefauthors}%
\unskip\
\newblock
\APACrefYearMonthDay{2008}{}{}.
\newblock
{\BBOQ}\APACrefatitle {Investigating {Ceiling} {Effects} in {Longitudinal} {Data} {Analysis}} {Investigating {Ceiling} {Effects} in {Longitudinal} {Data} {Analysis}}.{\BBCQ}
\newblock
\APACjournalVolNumPages{Multivariate Behavioral Research}{43}{3}{476--496}.
\newblock
\begin{APACrefDOI} \doi{10.1080/00273170802285941} \end{APACrefDOI}
\PrintBackRefs{\CurrentBib}

\bibitem [\protect \citeauthoryear {%
Weng%
}{%
Weng%
}{%
{\protect \APACyear {1998}}%
}]{%
weng_scale_1998}
\APACinsertmetastar {%
weng_scale_1998}%
\begin{APACrefauthors}%
Weng, L\BPBI J.%
\end{APACrefauthors}%
\unskip\
\newblock
\APACrefYearMonthDay{1998}{}{}.
\newblock
{\BBOQ}\APACrefatitle {Scale values of anchor labels in {Chinese} rating scales: {Responses} on frequency and agreement} {Scale values of anchor labels in {Chinese} rating scales: {Responses} on frequency and agreement}.{\BBCQ}
\newblock
\APACjournalVolNumPages{Chinese Journal of Psychology}{40}{}{73--86}.
\PrintBackRefs{\CurrentBib}

\bibitem [\protect \citeauthoryear {%
Williams%
, Morlock%
\BCBL {}\ \BBA {} Feltner%
}{%
Williams%
\ \protect \BOthers {.}}{%
{\protect \APACyear {2010}}%
}]{%
williams_psychometric_2010}
\APACinsertmetastar {%
williams_psychometric_2010}%
\begin{APACrefauthors}%
Williams, V\BPBI S.%
, Morlock, R\BPBI J.%
\BCBL {}\ \BBA {} Feltner, D.%
\end{APACrefauthors}%
\unskip\
\newblock
\APACrefYearMonthDay{2010}{}{}.
\newblock
{\BBOQ}\APACrefatitle {Psychometric evaluation of a visual analog scale for the assessment of anxiety} {Psychometric evaluation of a visual analog scale for the assessment of anxiety}.{\BBCQ}
\newblock
\APACjournalVolNumPages{Health and Quality of Life Outcomes}{8}{1}{57}.
\newblock
\begin{APACrefDOI} \doi{10.1186/1477-7525-8-57} \end{APACrefDOI}
\PrintBackRefs{\CurrentBib}

\bibitem [\protect \citeauthoryear {%
Wu%
\ \BBA {} Leung%
}{%
Wu%
\ \BBA {} Leung%
}{%
{\protect \APACyear {2017}}%
}]{%
wu_can_2017}
\APACinsertmetastar {%
wu_can_2017}%
\begin{APACrefauthors}%
Wu, H.%
\BCBT {}\ \BBA {} Leung, S\BHBI O.%
\end{APACrefauthors}%
\unskip\
\newblock
\APACrefYearMonthDay{2017}{}{}.
\newblock
{\BBOQ}\APACrefatitle {Can {Likert} {Scales} be {Treated} as {Interval} {Scales}?—{A} {Simulation} {Study}} {Can {Likert} {Scales} be {Treated} as {Interval} {Scales}?—{A} {Simulation} {Study}}.{\BBCQ}
\newblock
\APACjournalVolNumPages{Journal of Social Service Research}{43}{4}{527--532}.
\newblock
\begin{APACrefDOI} \doi{10.1080/01488376.2017.1329775} \end{APACrefDOI}
\PrintBackRefs{\CurrentBib}

\bibitem [\protect \citeauthoryear {%
Zealley%
\ \BBA {} Aitken%
}{%
Zealley%
\ \BBA {} Aitken%
}{%
{\protect \APACyear {1969}}%
}]{%
zealley_growing_1969}
\APACinsertmetastar {%
zealley_growing_1969}%
\begin{APACrefauthors}%
Zealley, A\BPBI K.%
\BCBT {}\ \BBA {} Aitken, R\BPBI C\BPBI B.%
\end{APACrefauthors}%
\unskip\
\newblock
\APACrefYearMonthDay{1969}{}{}.
\newblock
{\BBOQ}\APACrefatitle {A {Growing} {Edge} of {Measurement} of {Feelings} [ \textit{{Abridged}} ]: {Measurement} of {Mood}} {A {Growing} {Edge} of {Measurement} of {Feelings} [ \textit{{Abridged}} ]: {Measurement} of {Mood}}.{\BBCQ}
\newblock
\APACjournalVolNumPages{Proceedings of the Royal Society of Medicine}{62}{10}{993--996}.
\newblock
\begin{APACrefDOI} \doi{10.1177/003591576906201006} \end{APACrefDOI}
\PrintBackRefs{\CurrentBib}

\end{thebibliography}

\end{document}